\documentclass[reprint,groupedaddress,amsmath,amssymb,aps,pra,floatfix]{revtex4-1}

\newcommand{\bra}[1]{\langle #1\rvert}
\newcommand{\ket}[1]{\lvert #1\rangle}

\newcommand{\op}[2]{\ket{#1} \bra{#2}}
\newcommand{\exv}[1]{\langle #1\rangle}

\newcommand{\der}[1]{\frac{d #1}{dt}}
\newcommand{\pd}[1]{\frac{\partial #1}{\partial t}}
\newcommand{\rpd}[1]{\partial_t #1}

\usepackage{graphicx}% Include figure files
\usepackage{dcolumn}% Align table columns on decimal point
\usepackage{bm}% bold math
\usepackage{amsmath}
\usepackage{upgreek}
\usepackage{physics}
\usepackage{isomath}
\usepackage{placeins}

\begin{document}

\preprint{APS/123-QED}

\title{Speeding up particle slowing using shortcuts to adiabaticity}

\author{John P. Bartolotta}
\author{Jarrod T. Reilly}
\author{Murray J. Holland}

\affiliation{JILA and Department of Physics, University of Colorado, 440 UCB, Boulder, CO 80309, USA}

\date{\today}

\pacs{Valid PACS appear here}% PACS, the Physics and Astronomy
                             % Classification Scheme.
%\keywords{Suggested keywords}%Use showkeys class option if keyword
                              %display desired
\begin{abstract}
We propose a method for slowing particles by laser fields that potentially has the ability to generate large forces without the associated momentum diffusion that results from the random directions of spontaneously scattered photons. In this method, time-resolved laser pulses with periodically modified detunings address an ultranarrow electronic transition to reduce the particle momentum through repeated absorption and stimulated emission cycles. We implement a shortcut to adiabaticity approach that is based on Lewis-Riesenfeld invariant theory. This affords our scheme the advantages of adiabatic transfer, where there can be an intrinsic insensitivity to the precise strength and detuning characteristics of the applied field, with the advantages of rapid transfer that is necessary for obtaining a short slowing distance. For typical parameters of a thermal oven source that generates a particle beam with a central velocity on the order of meters per second, this could result in slowing the particles to near stationary in less than a millimeter. We compare the slowing scheme to widely-implemented slowing techniques that rely on radiation pressure forces and show the advantages that potentially arise when the excited state decay rate is small. Thus, this scheme is a particularly promising candidate to slow narrow-linewidth systems that lack closed cycling transitions, such as occurs in certain molecules. 
\end{abstract}

\maketitle

\begin{figure*}
\centerline{\includegraphics[width=0.8\linewidth]{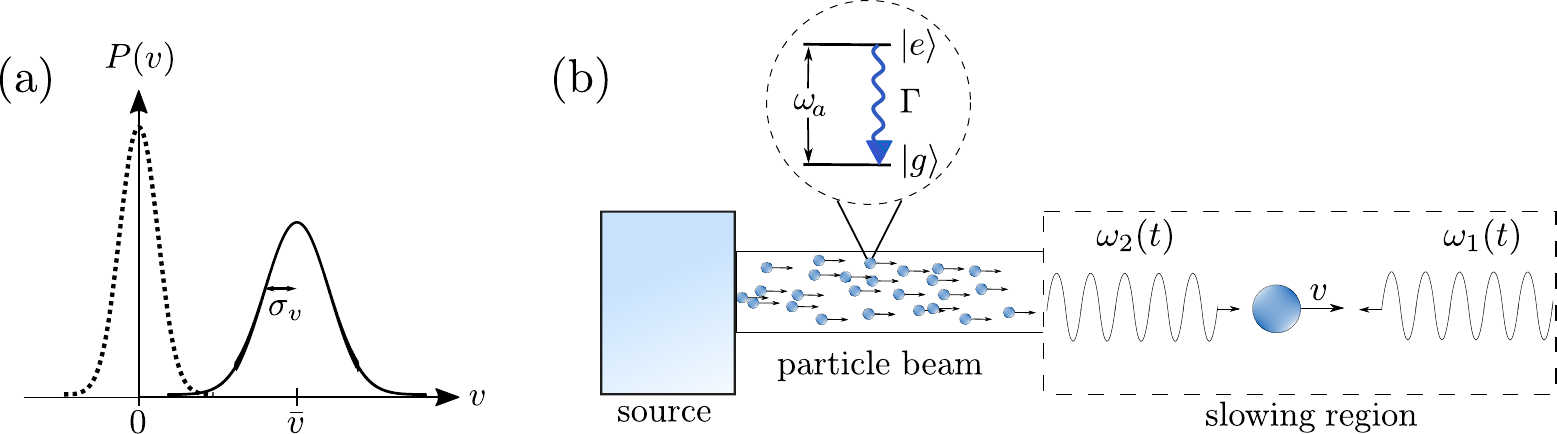}}
\caption{(a) Schematic of the normalized velocity distributions $P(v)$ of the particle beam before (solid) and after (dashed) the slowing process. We characterize the initial distribution by a Gaussian function with a standard deviation $\sigma_v$ and mean velocity $\bar v$. (b) Particles exit an atom or molecule source and are collimated into the slowing region. The spatial setup of the sequentially-pulsed counterpropagating lasers, which have time-dependent frequencies~$\omega_1(t)$ and $\omega_2(t)$, and a sample particle with velocity~$v$ in the laboratory frame are displayed in the slowing region. The circular inset shows the two-level internal structure of each particle.}
\label{expSchematic}
\end{figure*}

\section{Introduction}

For decades, laser cooling atoms and molecules to near absolute zero has been at the forefront of research into the interactions of light and matter \cite{chu,phillips,CT,Metcalf}. 
Ultracold ensembles provide testbeds for exploring fundamental physics~\cite{Safronova}, can create low-temperature superfluids such as Bose-Einstein and fermionic condensates~\cite{Cornell, Jin}, and can be used as platforms for quantum simulators~\cite{Kim_QS}.
Since atom and molecule sources typically produce particles with high velocities, it is usually necessary to first perform a precursor stage that removes a substantial fraction of the kinetic energy associated with the random thermal motion, as shown in Figure~\ref{expSchematic}(a).
%A specific example is sodium atoms produced in an oven at thermal velocities of a few hundred meters per second that need to be slowed to a few meters per second before being transferred into a magnetic and/or optical trap~\cite{SrSource,NaSlowing,ovens}. 
An assortment of particle slowing methods have been developed for this precursor stage, including Stark and Zeeman decelerators \cite{Narevicius,Akerman,Hogan,Petzold}, centrifuge decelerators \cite{Chervenkov}, electrostatic trapping methods \cite{Bethlem}, frequency-chirped laser slowing \cite{Truppe}, white light slowing \cite{Ye,Hemmerling}, and angled slowing \cite{Lunden}, among others.
Once slowed, the particles may possess sufficiently low kinetic energy that they can be efficiently loaded into a finite-depth electromagnetic trap \cite{Metcalf}, such as a magneto-optical trap (MOT), and then be cooled using light \cite{Stellmer,Urvoy}, cooled through evaporation that redistributes energy through two-body collisions \cite{Stuhl}, or sympathetically cooled with another species~\cite{Elliott}. 
%These basic steps have been the key to opening up the world of quantum gases to many modern applications of atomic and molecular physics in quantum science and engineering.

The ability to produce a large number of ultracold particles is made difficult by practical shortcomings of slowing methods, such as a large slowing distance that requires significant physical space, or substantial spread in the final velocities of the particles. Furthermore,
the main hindrance to slowing particles that lack closed cycling transitions is that there may be leakage of population to dark electronic states that are not coupled with the fields that perform the laser cooling and slowing, resulting in the loss of the particle from the system. Even if this does not occur, spontaneous emission of many photons creates momentum diffusion due to the random emission direction, and this results in heating and a finite limit on the achievable temperatures. 
%Recently, a laser cooling method called sawtooth-wave adiabatic passage (SWAP) cooling was experimentally demonstrated~\cite{SWAPExperiment} that may mitigate some of these shortcomings. The method was studied theoretically \cite{SWAPTheory}, applied to a Raman transition between hyperfine states~\cite{SWAPRaman}, and applied to a MOT~\cite{SWAP_MOT_theory,Snigirev,muniz}. 
These issues make methods that increase slowing forces and minimize the number of scattering events through enhanced control of tailored coherent dynamics, such as sawtooth-wave adiabatic passage (SWAP) cooling \cite{SWAPExperiment,SWAPTheory,SWAPRaman,SWAP_MOT_theory,Snigirev,muniz}, the Allen-Eberly scheme \cite{AE}, stimulated Raman adiabatic passage (STIRAP), the adiabatic passage force, and the bichromatic force \cite{Voitsekhovich,SalomonBCF,MetcalfBCF,ARP1,ARP2,MetcalfColloquium} enticing candidates to consider for particle slowing. However, one concern is that in order to satisfy an intrinsic adiabaticity condition, the time evolution should typically be slow and this could result in a long stopping time and associated large stopping distance.

In order to speed up adiabatic processes while still achieving substantial population transfer, there has been a growing interest in so-called ``shortcuts to adiabaticity"~\cite{Review}, i.e., processes that are able to transform systems from an initial quantum state to the same final quantum state as an ideal adiabatic process but in much less time. In this paper, we present a fast, simple, and robust particle slowing scheme that employs one such shortcut method: inverse engineering based on Lewis-Riesenfeld invariants (LRI) \cite{LRI,LRIandTQD,LRI_PRL,Lai,Chen_2014,LRIFourLevel}.
%A variety of techniques have been proposed and experimentally demonstrated: 
%Berry's transitionless quantum driving (TQD) algorithm \cite{Berry,SHAPE,Bason}, inverse engineering based on Lewis-Riesenfeld invariants \cite{LRI,LRIandTQD,LRI_PRL,Lai,Chen_2014,LRIFourLevel}, optimal control bang-bang type \cite{Viola,Caneva,Han,Rey}, fast-forward techniques for Schr\"odinger \cite{Nakamura,Masuda,Torrontegui} and Dirac dynamics \cite{Deffner_2015}, ``environment" assisted methods \cite{Rice}, fast quasiadiabatic dynamics (FAQUAD) \cite{Gillet}, and using the properties of Lie algebras \cite{Sarandy,Martinez-Garaot,Torrontegui_2014,wimberger}.
%So far, these adiabatic shortcut techniques have been employed in frictionless cooling in harmonic traps \cite{LRI_PRL}, in Penning traps \cite{Rey}, in cavity quantum electrodynamics \cite{Chen_2014}, in suppressing pair production \cite{Deffner_2015}, in atom interferometry \cite{Du}, in implementing the Allen-Eberly scheme \cite{SHAPE}, in STIRAP demonstrations \cite{SHAPE,Rice,Li,Mortensen}, in quantum simulators \cite{Han,gueryodelin}, in quantum computing \cite{Takahashi}, and even in quantum gaming~\cite{Sels}.
The slowing scheme involves driving sped-up transitions from a stable ground state to an excited state and back using counter-propagating, pulsed lasers with intensity and detuning profiles prescribed by the LRI shortcut method, as shown schematically in Figure~\ref{expSchematic}(b). By applying this protocol many times, the particle can in principle be subject to the impulse of many photon momenta without emitting spontaneous photons. In Section~\ref{utility}, we further motivate the use of our slowing scheme. In Section~\ref{LRITheory}, we outline the general theory of LRI-based inverse engineering. In Section~\ref{model}, we present our theoretical model and derive the laser intensity and detuning profiles that result from the LRI shortcut method.
%Repeating many cycles of these sped-up transitions will provide our scheme the advantages of adiabatic slowing but with a reduced slowing time and slowing distance. 
In Section~\ref{forcesAndSlowing}, we study the resulting classical forces, slowing times, and slowing distances, demonstrating that we are able to generate higher forces than what is achieved with radiation pressure as employed in methods such as Zeeman slowing.
In Section~\ref{SlowingExample}, we simulate particle slowing dynamics by applying our method to various atomic and molecular species with narrow linewidth transitions.
In Section~\ref{robustness}, we study our slowing scheme's robustness to various systematic errors that one may encounter when implementing the protocol in an experimental setting, and compare these results to slowing with $\pi$-pulse transitions and SWAP slowing.

\section{Motivation for the speed-up protocol}
\label{utility}

%In this section, we provide motivation for implementation of our protocol. 
The concept of coherent transfer between quantum states is a ubiquitous component of most quantum control techniques. The most basic way to achieve such a transfer between two electronic quantum states is to  
%introduce a coupling between the states with strength $\Omega$ such that the Hamiltonian can be expressed in the interaction picture as
%\begin{equation}
%    \hat H = \frac{\hbar \Omega}{2}
%    \left(
%    \ket{1}\bra{2} + \ket{2} \bra{1}
%    \right),
%\end{equation}
%where $\hbar$ is the reduced Planck constant and without loss of generality we have taken $\Omega$ to be real. In the case of light-matter interactions, this coupling is typically achieved by applying a coherent light source to a quantum particle, such as an atom or molecule, with a frequency $\omega$ that is resonant with the energy difference $E_2-E_1$ of the quantum states. This interaction results in population transfer between $\ket{1}$ and $\ket{2}$ that occurs at a frequency $\Omega/2$. In other words, if the quantum particle begins the interaction in the state $\ket{1}$, then it will be completely transferred to the state $\ket{2}$ at any time $t$ such that
%\begin{equation}
%    \Omega t = n\pi,
%\end{equation}
%where $n$ is any odd integer, and back to $\ket{1}$ when $n$ is an even integer. 
apply a resonant, coherent light pulse for a time $t=\pi/\Omega$, where $\Omega$ is the Rabi frequency. While this Rabi-flopping ``$\pi$-pulse" method completely transfers population, at least in principle, it is not robust to small errors in $\Omega$, the frequency $\omega$ of the light source, or coupling to other states outside of the two-level manifold.
%A further detrimental effect arises when applying this method to quantum particles with motion due to the associated Doppler shift $kv$ of the laser light experienced by the particle, where $v$ is the velocity of the particle as measured in the laboratory frame and $k=2\pi /\lambda$ is the wavenumber of the transition with wavelength $\lambda$. This shifts the laser light away from resonance with the particle, which changes both the rate and maximum amount of population transfer between the states. 
%This issue can be remedied by tuning the laser light back into resonance with the particle, but this is of limited utility when the goal is to transfer population in many particles that possess a distribution of velocities. 
Moreover, this method is of limited utility when the goal is to transfer population in many particles that possess a distribution of velocities. 

A robust solution to these problems is to chirp the laser frequency through the resonance, as was shown by Landau and Zener in their theory of adiabatic passage~\cite{zener}. However, such a method is limited in the sense that the laser chirping must be adiabatic, i.e., sufficiently slow compared to the strength of the coupling~$\Omega$.
%More specifically, they showed that the population of state $\ket{2}$, $P_2$, after the laser is chirped from a detuning of minus infinity to positive infinity, is
%\begin{equation}
%\label{LZprob}
%    P_2 = 1 - \exp
%    \left(
%    - \frac{\pi}{2} \frac{\Omega^2}{\alpha}
%    \right),
%\end{equation}
%where $\alpha$ is the frequency chirping rate in rad/s$^2$. 
%Thus, if the experiment is operated in the limit $\Omega^2 \gg \alpha$, then $P_2 \approx 1$, regardless of the particle's initial velocity, a phenomenon known as adiabatic passage. 
%The requirement of infinite time can be relaxed if the laser frequency is adiabatically chirped over a frequency range larger than $2 \Omega$ in the moving frame of every particle, as can be seen in a dressed state picture.
%The velocity-independent feature of adiabatic passage motivated the development of SWAP cooling \cite{SWAPExperiment,SWAPTheory,SWAP_MOT_theory,SWAPRaman,muniz,Snigirev}, which utilizes periodic frequency chirps to coherently transfer a particle between quantum states in such a way that its momentum can be significantly reduced. A natural extension of SWAP cooling is to employ its ability to apply large coherent forces in the process of particle slowing. However, there are several issues that arise in this venture. The obvious issue is the requirement of adiabaticity, which limits the timescale for population transfer and can lead to undesirably large slowing distances and times. 
Another potentially detrimental problem that arises from this slow evolution is the particle's tendency to relax from the quantum state with higher energy to the quantum state with lower energy through the process of spontaneous emission, which would disturb the coherent slowing process and introduce unwanted momentum diffusion. This motivates transferring as quickly as possible, and thereby satisfying the requirements, i.e., $\Omega \gg \Gamma$, for adiabatic {\it{rapid}} passage~\cite{Camparo_1984}. Furthermore, in the neighborhood of the resonance, there are significant oscillations in the populations that result from the precession of the Bloch vector as it travels along the Bloch sphere~\cite{feynman_vernon_hellwarth_1957,vitanov}, which can complicate the amount of population transfer. 
%A detailed investigation of these oscillations and other important timescales in Landau-Zener transition theory is presented in~\cite{vitanov}.

As we will show, our derivation of an alternative scheme for coherent transfer, by use of LRI theory, ameliorates every issue we have discussed thus far. We drive a different path along the Bloch sphere (see Section~\ref{bloch}) in order to speed up the transfer time and remove the high frequency population oscillations while utilizing the same amount of laser power. But perhaps the most appealing feature of our scheme is the ability to \emph{exactly} transfer population between quantum states in a finite time, which in principle permits high-fidelity slowing of a particle with arbitrarily high momentum to near rest.

\section{Lewis-Riesenfeld Invariant-based inverse engineering}
\label{LRITheory}

In this section, we present a brief overview of the LRI shortcut method. This approach allows us to derive specific laser intensity and detuning profiles that can be used to generate fast dynamics with the same results as adiabatic processes.

A dynamical invariant $\hat I(t)$ is a Hermitian operator
with a time-independent expectation value, i.e.,
\begin{equation}
\label{invariant}
\exv{\hat I} = \bra{\Psi(t)} \hat I(t) \ket{\Psi(t)} =
\text{const},
\end{equation}
where $\ket{\Psi(t)}$ is the state vector evolved by the Hamiltonian $\hat H(t)$.
It can be shown that the states $\ket{\psi_n(t)}$, defined by a gauge transformation
\begin{equation}
\label{gauge}
    \ket{\psi_n(t)} =
    e^{i \alpha_n(t)} \ket{\phi_n(t)}
\end{equation}
 of the eigenbasis $\ket{\phi_n(t)}$ of $\hat{I}(t)$, are each a solution to the time-dependent Schr\"odinger equation.
In Eq.~\eqref{gauge}, the ``Lewis-Riesenfeld phases'' $\alpha_n(t)$ are defined as 
\begin{equation} \label{LRPhases}
\alpha_n(t) = \frac{1}{\hbar} \int^t_{t_0} \bra{\phi_n(t')} i \hbar \frac{\partial}{\partial t'} - \hat H (t') \ket{\phi_n(t')} dt',
\end{equation}
where $t_0$ is some initial reference time and $\hbar$ is the reduced Planck constant.
It follows that a general solution $\ket{\Psi}$ to the Schr\"odinger equation can be decomposed as
\begin{equation} \label{GeneralSolution}
\ket{\Psi(t)} = \sum_n c_n \ket{\psi_n (t)}
=\sum_n c_n e^{i \alpha_n (t)} \ket{\phi_n (t)} ,
\end{equation}
where $c_n$ are time-independent amplitudes \cite{LRI}. 
From Eq.~\eqref{GeneralSolution}, the unitary time-evolution operator in the invariant basis is
\begin{equation}
\label{eq:unitary}
\hat U(t) = \sum_n e^{i \alpha_n(t)} \op{\phi_n (t)}{\phi_n (t_0)} \text{,}
\end{equation}
which can be used to solve for the Hamiltonian:
\begin{equation}
\label{eq:HamiltonianUnitaryRel}
\hat H(t) = i \hbar \pd{\hat U (t)} \hat U^\dagger (t) \text{.}
\end{equation}

These results indicate numerous benefits for this method. In particular, if $\hat H(t)$ and $\hat I(t)$ are designed to commute at $t_0$ and some final time $t_f$, i.e.,
\begin{equation}
\label{eq:commutator}
    \left[ \hat{H}(t_0),\hat{I}(t_0) \right]=
    \left[ \hat{H}(t_f),\hat{I}(t_f) \right]= 0,
\end{equation} then the final state $\ket{\Psi(t_f)}$ will maintain the initial populations for each eigenstate \cite{Review}, and we therefore recover the results of an adiabatic process without the requirement of slow time-evolution. Also, by combining Eqns.~(\ref{eq:unitary}) and~(\ref{eq:HamiltonianUnitaryRel}), we can explicitly inverse-engineer the Hamiltonian in the invariant basis:
\begin{align} 
\hat H(t) = & -\hbar \sum_n \dot \alpha_n \op{\phi_n (t)}{\phi_n (t)} \notag \\
\label{LRIHamiltonian}
    & + i \hbar \sum_n \op{\rpd{\phi_n (t)}}{\phi_n (t)} \text{.}
\end{align}
Eqns.~\eqref{eq:HamiltonianUnitaryRel}-\eqref{LRIHamiltonian} define what is meant by invariant-based inverse engineering. Equating the Hamiltonian in the invariant basis to the original Hamiltonian creates a map between the physical parameters and the auxiliary parameters that define the invariant operator. In Section~\ref{InvariantsApproach}, We will use this procedure to derive the laser intensity and detuning profiles for each single-photon transition in our slowing scheme. 

\begin{figure*}
\centerline{\includegraphics[width=0.95\linewidth]{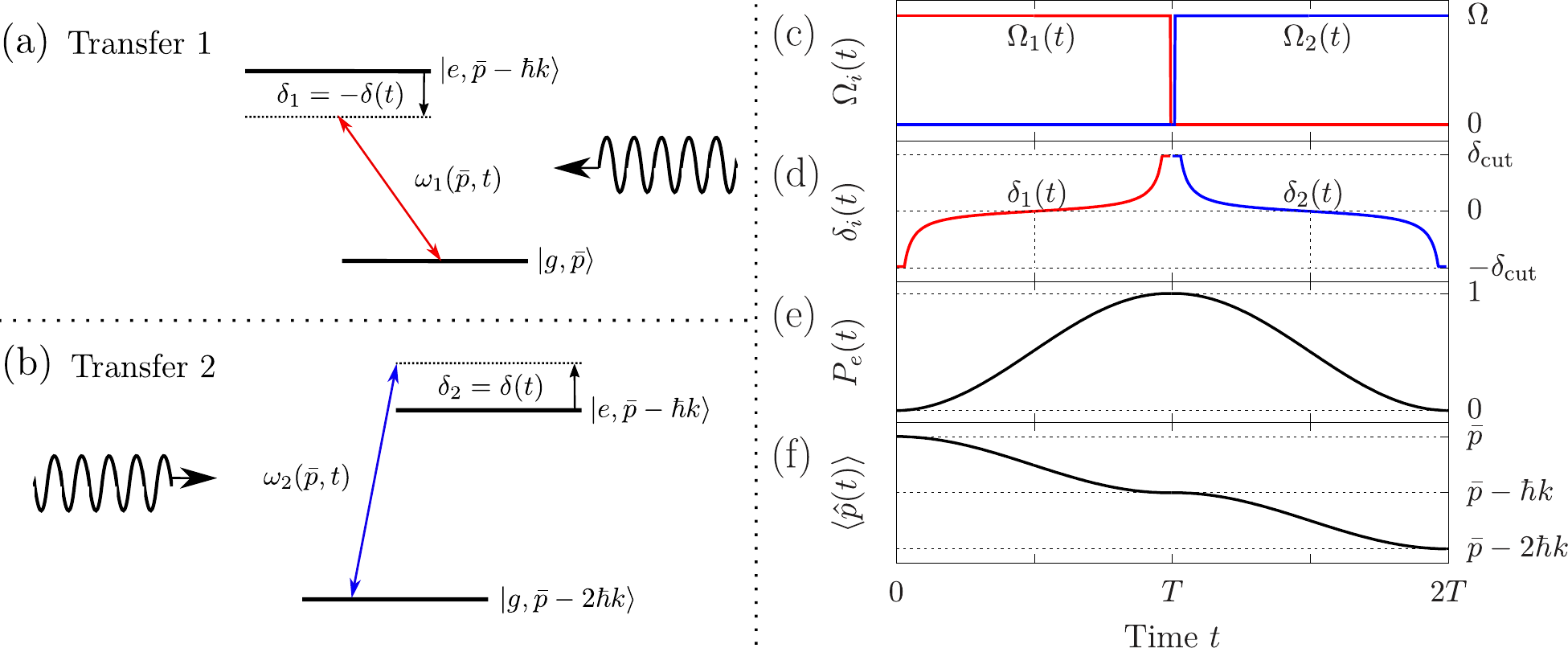}}
\caption{\textbf{Left}: frequency diagram of an isolated subset of states in the lab frame. The laser frequencies $\omega_i(\bar p,t)$ [see Eqns.~(\ref{eq:laserFreqs})] are dynamically updated according to the solution derived from the Lewis-Riesenfeld invariant shortcut method to promote quick, coherent transfer from (a) $\ket{g,\bar p}$ to $\ket{e,\bar p- \hbar k}$, followed by (b) $\ket{e,\bar p- \hbar k}$ to $\ket{g,\bar p- 2\hbar k}$. \textbf{Right}: experimental parameters and particle dynamics over a $\ket{\bar p} \rightarrow \ket{\bar p-2\hbar k}$ sequence of period $2T$. (c) Square-pulse Rabi frequency profiles $\Omega_1(t)$ (red) and $\Omega_2(t)$ (blue) with amplitude $\Omega$ [see Eq.~\eqref{eq:Omega}]. (d) Lewis-Riesenfeld detuning profile $\delta(t)$ [see Eq.~\eqref{eq:LRIdetuning}] with cutoff frequency $\pm \delta_\text{cut}$ [see Eq.~\eqref{cutoff}] for each laser. (e) Ideal excited state fraction $P_e$ dynamics. (f) Ideal average momentum $\langle \hat p\rangle $ dynamics. Parameters are: $T = 0.032/\omega_r, \, \Omega = 100\omega_r, \, \delta_\text{cut}=250\omega_r, \, \Gamma = 0, \, \bar p = 100 \hbar k$, and $\beta = 0.85 \, \pi /2$.}
\label{scheme}
\end{figure*}

\section{Model} \label{model}
We consider the experimental setup depicted in Fig.~\ref{expSchematic}(b).
As shown in the circular inset, we consider each particle to possess two internal electronic states labeled $\ket{g}$ and $\ket{e}$ with an energy separation of $\hbar \omega_a$, where $\omega_a$ is a frequency assumed to be in the optical domain. The excited state $\ket{e}$ can decay to the ground state $\ket{g}$ at a rate given by the natural linewidth $\Gamma$. We track motion along one dimension, for which the particle has position and momentum operators 
$\hat{z}$ and $\hat{p}$. 
%Since this slowing scheme is essentially translationally invariant, we choose to represent the physics in the momentum basis.
A thermal beam of particles exits an atom or molecule source (e.g., an oven) at a high average velocity $\bar v$ and potentially large velocity spread~$\sigma_v$. The goal is to quickly remove kinetic energy from the particles such that an appreciable fraction of the final distribution is centered on zero velocity [see Fig.~\ref{expSchematic}(a)]. In order to achieve this goal, the particle beam sequentially interacts with pulsed slowing lasers aligned parallel to the beam axis upon entering the slowing region. The essential dynamics are displayed in Fig.~\ref{scheme} and are as follows: The counterpropagating laser (laser 1) is switched on, and the particle absorbs a photon and transits to $\ket{e}$. Then, laser 1 is switched off as the copropagating laser (laser 2) is switched on, and the particle emits a photon by stimulated emission, transitioning back to $\ket{g}$. By conservation of momentum, the particle has experienced an impulse of $2 \hbar k$ by the end of the sequence. By repeating many times, we can remove many photon momenta from the particle without the emission of spontaneous photons, contingent that the operation occurs quickly compared to any decoherence processes.

We set the Rabi frequency of each laser $\Omega_i(t)$ (with $i =1,2$) to follow a square pulse temporal profile. The peak value~$\Omega$ [see Fig.~\ref{scheme}(c)] will be derived from the LRI shortcut method. This choice of waveform removes the requirement of exactly aligning periodic Rabi frequency and laser detuning profiles in time as is needed in other protocols \cite{eberly,ARP1,ARP2}, and thus makes the process more robust. We parameterize the instantaneous frequencies $\omega_1(t)$ and $\omega_2(t)$ of each laser field in the form
\begin{equation}
\label{eq:laserFreqs}
    \begin{aligned} \omega_1(\bar p,t) &= \omega_a - \delta(t) - k\bar v + \omega_r \\
    \omega_2(\bar p,t) &= \omega_a + \delta(t) + k\bar v - 3\omega_r,
    \end{aligned}
\end{equation}
so that, after accounting for the mean particle velocity $\bar v=\bar p/m$, each laser is resonant with the transitions displayed in Fig.~\ref{scheme}(a) and Fig.~\ref{scheme}(b) when the detuning $\delta(t)$ is zero. We will derive an explicit form for $\delta(t)$ from the LRI shortcut method. Here, $\omega_r \equiv \hbar k^2/2m$ is the recoil frequency of the transition, i.e. $\hbar\omega_r$ is the kinetic energy obtained by an atom at rest from the absorption of a single photon, and $m$ is the particle mass. Also, we have made the typical approximation that the wavenumber $k_i(t)$ of each laser is equal to the constant $k$ at all times, even though the frequency is varying. It should be pointed out that although we have chosen specific signs for $\delta(t)$ in each laser frequency $\omega_i(\bar p, t)$ [see Eq.~\eqref{eq:laserFreqs}], these are not the only conventions that can be employed to result in slowing.

\subsection{System dynamics}

When applied to a system with momentum $\bar p = m \bar v$, the Hamiltonian takes the general form
\begin{equation}
    \hat{H}(\bar p,t) = \hat{H}_\text{self} + \hat{H}_\text{int}(\bar p,t),
\end{equation}
where
\begin{equation}
\label{eq:freeHam}
\hat{H}_\text{self}=\frac{\hat{p}^2}{2m} + \frac{\hbar \omega_a}{2} \hat \sigma^z
\end{equation}
is the particle's free evolution Hamiltonian. The quantity $\hat \sigma^z \equiv \ket{e} \bra{e} - \ket{g} \bra{g}$ is the usual Pauli spin matrix. Under the dipole and rotating wave approximations, the particle-field interaction Hamiltonian $\hat{H}_\text{int}(\bar p,t)$ in the Schr\"odinger picture is given by
\begin{eqnarray}
\hat{H}_\text{int}(\bar p,t)&=&
\frac{\hbar}{2}\,\hat{\sigma}^-\Bigl[
\Omega_1(t)
e^{i[k\hat{z}+\eta_1(\bar p,t)]}\nonumber\\
&&{}+\Omega_2(t)e^{-i[k\hat{z}-\eta_2(\bar p,t)]}\Bigr]+\text{h.c.}
\label{eq:intHam}
\end{eqnarray}
where $\hat{\sigma}^- \equiv \ket{g}\bra{e}$ is the lowering operator. The quantity $\eta_i(\bar p,t)$ is the accumulated phase from an initial time $t_0$ for each laser field [see Eq.~\eqref{eq:laserFreqs}]:
\begin{equation}
\label{eq:accumulatedPhase}
\eta_i(\bar p,t) \equiv \int_{t_0}^t \omega_i(\bar p,t') dt' \text{.}
\end{equation}

In order to capture the relevant physics in the first half of the sequence, we omit the second term in Eq.~(\ref{eq:intHam}) (since laser 2 is off) and transform into the interaction picture defined by the Hamiltonian 
\begin{equation}
    \hat H_0(t) = 
    \frac{\hat p^2}{2m} 
    +  \frac{\hbar}{2} 
    \left[\omega_a-\delta(t)\right] \hat \sigma^z.
\end{equation}
The resulting interaction picture Hamiltonian $\hat{H}_1(\bar p,t)$ is
\begin{equation}
\label{eq:HInt}
\begin{aligned}
    & \hat{H}_1(\bar p,t) = \frac{\hbar \delta(t)}{2}
    \hat \sigma^z +\\
    &\frac{\hbar \Omega_1(t)}{2}
    \Big(
    \hat{\sigma}^-
    \exp\left(
    i\left[
    k\hat{z}+k(\hat{v} - \bar v)t + \omega_r t
    \right]\right)
    + \text{h.c.}
    \Big),
\end{aligned}
\end{equation}
where $\hat v= \hat p /m$ is the velocity operator. The interaction picture Hamiltonian for the second half $\hat H_2(\bar p,t)$ is found by the substitutions $1 \rightarrow 2$, $\delta(t) \rightarrow - \delta(t),$ $k \rightarrow -k$, and $\omega_r \rightarrow -3 \omega_r$. After the complete evolution with $\hat H_1(\bar p,t)$ followed by  $\hat H_2(\bar p,t)$, the momentum $\bar p$ is updated to $\bar p - 2 \hbar k$ for the laser detunings [i.e., see Eq.~(\ref{eq:laserFreqs})] under the assumption that the particle momentum has been changed accordingly, and the process repeated.

We incorporate the effects of incoherent dynamics due to spontaneous emission and its associated recoil by evolving the density matrix operator $\hat{\rho}$ of the system under the quantum master equation,
\begin{equation} \label{MasterEq}
\der{\hat \rho} = \frac{1}{i \hbar} \left[ \hat H, \hat \rho \right] + \hat{\mathcal{L}} \left( \hat \rho \right) \text{,}
\end{equation}
where the Lindblad superoperator $\hat{\mathcal{L}}$ is
\begin{equation}
	\begin{aligned}
\hat{\mathcal{L}} (\hat \rho) &= - \frac{\Gamma}{2} \left[ \hat \sigma^+ \hat \sigma^- \hat \rho + \hat \rho \hat \sigma^+ \hat \sigma^- - \frac25 \left( 3\, \hat \sigma^- \hat \rho \hat \sigma^+ \right. \right. \\
&+ \left. \left.  e^{ik \hat z} \hat \sigma^- \hat \rho \hat \sigma^+ e^{-ik \hat z} +  e^{-ik \hat z} \hat \sigma^- \hat \rho \hat \sigma^+ e^{ik \hat z} \right) \right] \text{.}
	\end{aligned}
\end{equation}
Here, $\Gamma$ is the optical transition linewidth, and, for convenience, we have approximated the continuous dipole radiation pattern as allowing only the three discrete recoil possibilities that correspond to whole intervals of photon quanta. This implies possible impulses $\Delta p$ of $-\hbar k, 0$, and $\hbar k$ along the slowing axis with associated probabilities of $\frac{1}{5}, \frac{3}{5}$, and $\frac{1}{5}$, respectively \cite{Molmer}. 

\subsection{Eigensystem in a momentum subspace}

Our next task is to find appropriate forms for the Rabi frequencies $\Omega_i(t)$ and detuning $\delta(t)$ by following the shortcut protocol. In order to do this, we will perform analytical calculations for transfer 1 in a small, isolated subset $W(\bar p)$ of the full composite Hilbert space:
\begin{equation}
\label{eq:subset}
W(\bar p)=\{\ket{g,\bar p}, \ket{e,\bar p-\hbar k}\} = \{\ket{G},\ket{E}\}.
\end{equation}
The subset 
\begin{equation}
\label{eq:subset2}
W'(\bar p)=\{\ket{e,\bar p-\hbar k},\ket{g,\bar p-2 \hbar k}\}
\end{equation}
is used to calculate the corresponding quantities for transfer 2 [see Fig.~\ref{scheme}(b)], but we do not explicitly show the derivation here as it follows in a straightforward manner.
We expect the resulting shortcuts to be useful for application to the entire Hilbert space because the dynamics that result from the interaction with each traveling wave are relatively simple, i.e., there are no multiphoton resonances that would cause population transfer to other quantum states~\cite{dopplerons}. In Section~\ref{SlowingExample}, we apply the resulting Rabi frequencies and laser frequencies from this calculation to the entire system in a numerical simulation.

The interaction Hamiltonian $\hat H_1(t)$ written in the subspace $W(\bar p)$ is given by
\begin{equation}
\label{eq:effectiveHam}
    \hat{H}_1^{(W)}(t) = 
    \frac{\hbar \delta(t)}{2} 
    \hat \sigma_W^z
    +
    \frac{\hbar \Omega_1(t)}{2} 
    \hat \sigma_W^x
\end{equation}
where $\hat \sigma_W^z \equiv \op{E}{E} - \op{G}{G}$ and $\hat \sigma_W^x \equiv \ket{G}\bra{E} +\ket{E}\bra{G}$. As we will show in subsection~\ref{InvariantsApproach}, we cast the invariant eigenvectors in a similar form as the eigenvectors of $\hat H_1^{(W)}(t)$ to apply the LRI shortcut method. The eigenvalues of $\hat H_1^{(W)}(t)$ are
\begin{equation}
E_{\pm} (t) = \pm \frac{\hbar \tilde \Omega_1 (t)}{2} \text{,}
\end{equation}
where $\tilde \Omega_1(t) \equiv \sqrt{\delta(t)^2 + \Omega_1(t)^2}$ is the generalized Rabi frequency.
The corresponding instantaneous eigenvectors are
\begin{equation}
\begin{aligned}
\ket{+} & = \cos \left( \frac{\chi}{2} \right) \ket{E} + \sin \left( \frac{\chi}{2} \right) \ket{G} \text{,} \\
\label{+-}
\ket{-} & = \sin \left( \frac{\chi}{2} \right) \ket{E} - \cos \left( \frac{\chi}{2} \right) \ket{G} \text{,}
\end{aligned}
\end{equation}
with the mixing angle defined by $ \cos\chi\equiv \delta(t)/\tilde \Omega_1(t)$. 

In general, if a system begins in an eigenstate of the Hamiltonian and the time evolution obeys the conditions of the adiabatic approximation, the state of the particle will simply adiabatically follow, only picking up an inconsequential global phase.
Therefore, if one chooses experimental parameters such that the particle is initialized in one of the dressed states and the dressed state adiabatically evolves from $\ket{G}$ to $\ket{E}$ (or vice versa), then we can change the particle momentum by $ \hbar k$.
However, these adiabatic states must evolve slowly to satisfy the adiabatic condition that arises from Landau-Zener theory.
We shall now speed up this adiabatic passage by constructing a driving Hamiltonian based on the LRI shortcut method.

\subsection{Deriving detuning profiles from the auxiliary equations} \label{InvariantsApproach}

We are now in a position to design an invariant operator in order to end up with the desired final populations using the formalism established in Section~\ref{LRITheory}. 
We provide the details of this process in Appendix~\ref{invariantConstruction} and give the important results here.

We parameterize the eigenvectors of the invariant $\hat I (t)$ in parallel to Eq.~\eqref{+-}:
\begin{equation}
\begin{aligned}
\ket{\phi_+ (t)} &= \cos \left( \frac{\gamma}{2} \right) e^{i \beta} \ket{E} + \sin \left( \frac{\gamma}{2} \right) \ket{G} \text{,} \\
 \label{eigenvectors}
\ket{\phi_- (t)} &= \sin \left( \frac{\gamma}{2} \right) \ket{E} - \cos \left( \frac{\gamma}{2} \right) e^{-i \beta} \ket{G} \text{,}
\end{aligned}
\end{equation}
where $\gamma = \gamma (t)$ and $\beta = \text{const}$ are auxiliary angles. We have introduced the unitary phase $e^{i \beta}$ parameterized by $\beta$ as an additional degree of freedom to explore with the shortcut protocol [see Section~\ref{bloch}]. From Eqns.~\eqref{LRPhases} and~\eqref{LRIHamiltonian}, the auxiliary equations are
\begin{align} \label{AuxEq1}
\dot \gamma &= \Omega_1(t) \sin \beta {,} \\
\label{AuxEq2}
\delta(t) & = \Omega_1(t) \cot \gamma \cos \beta \text{,}
\end{align}
which must satisfy the boundary conditions
\begin{equation}
\label{eq:gammaBoundary}
    \gamma(t_0) = 0, \qquad
    \gamma(t_f) = \pi
\end{equation} 
at the initial and final times $t_0$ and $t_f$. 

Together Eqns.~\eqref{AuxEq1}-\eqref{eq:gammaBoundary} directly determine the experimental parameters $\delta (t)$ and $\Omega_1(t)$. In the general case of $\Omega_1(t)\ne\text{const}$, we would be faced with the task of choosing, from an infinite set of Hamiltonians, a particular form of $\gamma (t)$ and $\beta (t)$ that satisfies Eqns.~\eqref{AuxEq1}-\eqref{eq:gammaBoundary} \cite{LRIandTQD}. However, we will now simplify the discussion by restricting the solution to the form $\Omega_1(t) = \Omega=\text{const}$, whereby the auxiliary equations~(\ref{AuxEq1}) and~(\ref{AuxEq2}) can be solved analytically, which we now show. Let us label the time required to perform the shortcut as $T$. Defining $t_0=0$ (and therefore $t_f=T$), the auxiliary equation for $\gamma$ using $\gamma(0)=0$ yields
\begin{equation}
\label{eq:gammaFirstSol}
    \gamma(t) = (\Omega \sin \beta) t.
\end{equation}
The other boundary condition $\gamma(T)=\pi$ [see Eq.~(\ref{eq:gammaBoundary})] requires that
\begin{equation}
\label{eq:Omega}
    \pi = \Omega T \sin \beta.
\end{equation}
Combining Eqns.~(\ref{eq:gammaFirstSol}) and~(\ref{eq:Omega}) and substituting the result into Eq.~(\ref{AuxEq2}), we find that
\begin{equation}
\label{eq:LRIdetuning}
    \delta(t) = \frac{\pi \cot \beta}{T} 
    \cot 
    \left(
    \frac{\pi t}{T}
    \right).
\end{equation}
We provide a plot of $\delta(t)$ for a particular choice of $\beta$ and $T$ in Fig.~\ref{scheme}(d).

We have now found a particular Hamiltonian that will drive the transition $\ket{G} \rightarrow \ket{E}$ with dynamics that need not be adiabatic. The analogous solution can be used for laser 2 to drive the transition $\ket{e,\bar p-\hbar k} \rightarrow \ket{g,\bar p- 2 \hbar k}$ using precisely the same theoretical formalism. The method may be optimized further with respect to different cost functions (e.g., see Ref.~\cite{Stefanatos}).

\subsection{Bloch sphere trajectories}
\label{bloch}

\begin{figure}
\centerline{\includegraphics[width=0.75\linewidth]{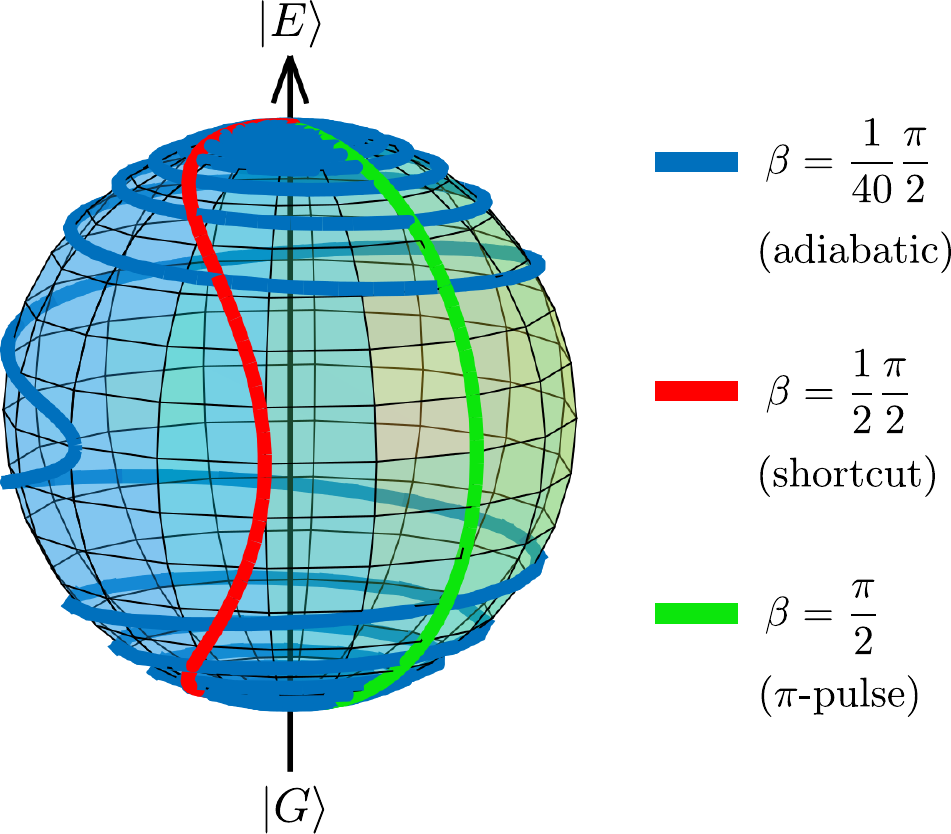}}
\caption{Bloch sphere trajectories (thick curves) between the initial state $\ket{G}$ and final state $\ket{E}$ parameterized by the auxiliary angle $\beta$ [see Eq.~\eqref{eigenvectors}] in the free-energy interaction picture [see Eq.~\eqref{blochHam}]. We set the Rabi frequency $\Omega$ to be equal for all trajectories, and the slowing periods $T$ are given by Eq.~\eqref{eq:Omega}. The cutoff detunings for $\beta=\tfrac{1}{40}\tfrac{\pi}{2}$ and $\beta=\tfrac{1}{2}\tfrac{\pi}{2}$ are $\delta_\text{cut}=318 \Omega$ and $\delta_\text{cut}=225 \Omega$, respectively [see Eq.~\eqref{cutoff}].}
\label{blochFig}
\end{figure}

Different choices of the auxiliary angle $\beta$ [see Eq.~\eqref{eigenvectors}] can result in very distinct dynamics as the particle is transferred from $\ket{G}$ to $\ket{E}$, which can be seen by investigating the associated Bloch sphere trajectories. For this purpose, we work in the time-independent interaction picture defined by the self-energy of the particle [see Eq.~\eqref{eq:freeHam}], which results in the interaction picture Hamiltonian
\begin{equation}
\label{blochHam}
\hat H_I(t) = \frac{\hbar \Omega}{2}
    \left(
    e^{i \theta(t)} \ket{E}\bra{G} + \text{h.c.}
    \right)
\end{equation}
in the subspace $W(p)$ [see Eq.~\eqref{eq:subset}], where
\begin{equation}
    \theta(t) \equiv \int_{t_0}^t \delta(t') \, dt'
\end{equation}
is the accumulated phase of the detuning $\delta(t)$ derived from the LRI shortcut method [see Eq.~\eqref{eq:LRIdetuning}].

Bloch sphere trajectories for three choices of $\beta$ are displayed in Figure~\ref{blochFig}. For the sake of comparison, we chose to fix the Rabi frequency $\Omega$ across all trajectories so that the slowing periods $T$ are completely determined by $\beta$ [see Eq.~\eqref{eq:Omega}].
A choice of auxiliary angle satisfying $\beta = (r+\tfrac{1}{2}) \pi$ for integer $r$ minimizes the slowing period~$T$; this is simply resonant Rabi-flopping ($\delta(t)=0$). For $r=0$, i.e., a resonant $\pi$-pulse, the system follows the purely longitudinal Bloch sphere trajectory (green, vertical curve). This choice of $\beta$ can lead to experimental challenges, such as inefficient transfer due to sensitivity to errors in $\Omega$ (see Section~\ref{robustness}). On the other hand, a choice of auxiliary angle satisfying $\beta \approx s \pi$ for integer $s$ causes $T$ to tend to infinity, as in an adiabatic process. We approximate the choice $s=0$ by using the very small auxiliary angle $\beta = \tfrac{1}{40} \tfrac{\pi}{2}$, which results in a trajectory that precesses significantly (blue, spiral curve). This choice also comes with complications, such as a high chance of spontaneous emission and a large slowing distance. We take all of these considerations into account and choose from the range $0  \ll \beta < \pi/2$ in our slowing simulations in Section~\ref{SlowingExample}, which results in an intermediate trajectory (red, wavy curve).

%We can gain further intuition into the effects of $\beta$ by calculating the adiabaticity parameter at the center of the transfer process~\cite{SWAPTheory}:
%\begin{equation}
%\label{kappa}
%  \displaystyle \kappa \equiv \left. \frac{\Omega(t)^2}{\displaystyle \left( \frac{d \delta(t)}{dt}\right)}\right|_{\displaystyle t=T/2}.
%\end{equation}

\begin{table*}
\bgroup
\def\arraystretch{3}
\begin{tabular}{c||c|c|c|c|c}
 \multicolumn{6}{c}{} \\
 \hline
 slowing method & force $F$ & slowing time $\Delta t$ & slowing distance $\Delta x$ & scattering rate $R^s$ & number of scattered photons $N$\\
 \hline
 Lewis-Riesenfeld & $\displaystyle \frac{ \Omega \hbar k \sin \beta}{\pi}$ & $\displaystyle \frac{ \pi \zeta_0}{\Omega \sin \beta}$ & $\displaystyle \frac{\omega_r \zeta_0^2 }{2 \Omega \sin \beta} \lambda$ & $\displaystyle \frac{\Gamma}{2}$ & $\displaystyle \frac{\Gamma}{\Omega} \frac{\pi}{2 \sin \beta} \zeta_0$\\
  Radiation pressure& $\Gamma \hbar k \rho_\text{ee}$ & $\displaystyle \frac{\zeta_0}{\Gamma \rho_\text{ee}}$ & $\displaystyle  \frac{\omega_r \zeta_0^2}{2 \pi \rho_\text{ee}  \Gamma}\lambda$ & $\Gamma \rho_\text{ee}$ & $\zeta_0$\\
 \hline
\end{tabular}
\egroup
 \caption{Comparison of slowing dynamics between the shortcut slowing scheme based on Lewis-Riesenfeld invariants and radiation-pressure slowing. If $\Omega \sin \beta \gg \pi \Gamma \rho_\text{ee}$, the shortcut slowing scheme results in larger forces, shorter slowing times, shorter slowing distances, and fewer scattered photons than RP.}
 \label{dynamicsTable}
\end{table*}

\section{Slowing dynamics in classical phase space}
\label{forcesAndSlowing}

In order to interpret the usefulness of the dynamics under the LRI shortcut slowing scheme, we calculate the resulting classical force, slowing time, slowing distance, scattering rate, and number of scattered photons and compare the results to commonly-implemented processes that rely on radiation pressure (RP), such as a Zeeman or Stark decelerator. These quantities are derived below and collected in Table~\ref{dynamicsTable}.

We use Eq.~(\ref{eq:Omega}) in order to calculate the classical force $F_\text{LRI}$ exerted on the particle during each sweep:
\begin{equation}
\label{LRI_force}
F_\text{LRI} = \frac{\hbar k}{T} = \frac{ \Omega \hbar k \sin \beta}{\pi}.
\end{equation}
The classical force from RP is given by
\begin{equation} \label{rad_force}
F_\text{RP} = \Gamma \hbar k \rho_\text{ee}  \leq  \frac{\Gamma \hbar k}{2} \text{,}
\end{equation}
where $\rho_\text{ee}$ is the excited state fraction, and the inequality is saturated at infinite laser power \cite{Metcalf}.

Next, we consider the time required to transfer the particle to zero momentum.
We assume the particle begins the slowing process in the state $\ket{g, \bar p_0}$. It receives an impulse of $\hbar k$ against its motion after every sweep for a total time
\begin{equation}
\label{slowingTime}
    \Delta t_\text{LRI} = T \bar p_0 / \hbar k =  \frac{ \pi \zeta_0}{\Omega \sin \beta},
\end{equation}
where $\zeta_0 \equiv \bar p_0 /\hbar k$. Assuming that the force $F_\text{RP}$ remains constant, the slowing time for RP is
\begin{equation}
    \Delta t_\text{RP} = \frac{\zeta_0}{\Gamma \rho_\text{ee}},
\end{equation}
since approximately one scattered photon is required per slowing photon.

We now calculate the distance a particle in the final momentum wave packet travels in physical space throughout the slowing process in order to reach zero momentum.
We approximate $p(t)$ to be linear:
\begin{equation}
    p(t) \approx \bar p_0 - \hbar k t/T
\end{equation}
so we may write the slowing distance as 
\begin{equation}
\label{x_LRI_time}
\Delta x_\text{LRI} = \frac{p_\text{avg}}{m} \Delta t = \frac{\omega_r T\zeta_0^2 }{2 \pi} \lambda \text{,}
\end{equation}
where $p_\text{avg}=\bar p_0/2$ is the time-averaged momentum and $\lambda$ is the wavelength of the transition.
Because laser power is typically the limiting experimental factor, we use Eq.~\eqref{eq:Omega} to rewrite Eq.~\eqref{x_LRI_time} as
\begin{equation}
\label{xLRIOmega}
\Delta x_\text{LRI} = \frac{\omega_r \zeta_0^2 }{2 \Omega \sin \beta} \lambda.
\end{equation}
Using Eq.~\eqref{rad_force}, we can obtain a similar expression for the RP slowing distance:
\begin{equation}
\Delta x_\text{RP} = \frac{\omega_r \zeta_0^2}{2 \pi \rho_\text{ee}  \Gamma}\lambda.
\end{equation}
Comparing the LRI and RP results, we find that our scheme is able to exert higher forces, and therefore slow particles in less time and in a shorter distance, with fewer scattered photons when $\Omega \sin \beta \gg \pi \Gamma \rho_\text{ee}$. For choices of $\sin \beta$ and $\rho_\text{ee}$ on the order of unity, this amounts to the limit $\Omega \gg \Gamma$. 

In the LRI analysis, we have not incorporated any effects due to dissipation. Since the slowing mechanism is purely coherent in nature, it would be ideal if there were no spontaneous emission events. This can be achieved in several ways: the use of ultra-narrow linewidth transitions, applying the entire slowing protocol in a time shorter than the lifetime of the excited state (which typically requires extremely high laser power but also results in very small slowing distances), or applying the shortcut protocol on a Raman transition between internal ground states. The last method is beyond the scope of this work, but may extend the application of this protocol to other atomic and molecular species. Nevertheless, the scattering rate $R^s$ of each slowing technique is approximately  
\begin{equation}
    R^s_\text{LRI} \approx \frac{\Gamma}{2}; 
    \quad 
    R^s_\text{RP} \approx \Gamma \rho_\text{ee} \leq \frac{\Gamma}{2},
\end{equation}
since the particles in the LRI scheme are in the excited state for roughly half of the time, suggesting that the scattering rates are on the same order. However, the expected number of scattered photons, $N= R^s \Delta t$, for each technique is
\begin{equation}
\label{numberSE}
    N_\text{LRI} \approx \frac{\Gamma}{\Omega} \frac{\pi}{2 \sin \beta} \zeta_0; 
    \quad 
    N_\text{RP} \approx \zeta_0,
\end{equation}
which means that the LRI scheme generates fewer scattered photons when $\Omega \sin \beta \gg \Gamma$.

\section{Slowing Examples} \label{SlowingExample}

By repeatedly driving the transition $\ket{g,\bar p} \rightarrow \ket{e,\bar p- \hbar k}$ followed by $\ket{e,\bar p-\hbar k} \rightarrow \ket{g,\bar p- 2\hbar k}$ according to our shortcut solution, we now show that we are able to slow a significant number of particles when the full composite Hilbert space is taken into account. After each pair of transitions, we update the momentum $\bar p$ to $\bar p - 2 \hbar k$ in the laser frequencies [see Eq.~\eqref{eq:laserFreqs}], as we assume that the initial population in $\ket{g,\bar p}$ has moved to $\ket{g,\bar p-2\hbar k}$. Note that this procedure results in the slowing of a pulse, as opposed to a steady-state ensemble, of particles. We assume that the jump in laser frequency at the end of each ramp is perfectly diabatic. Since there is very little population transfer when $\delta(t) \gg \Omega$, we do not track the detuning profile's asymptotic behavior, as this can cause numerical errors. 
We thus take $\delta (t)$ to be equal to a constant $\pm \delta_\text{cut}$ in these regions, defined by
\begin{equation}
\label{cutoff}
    \delta (t<t_0 + t_\text{cut}) 
    = - \delta (t> t_f-t_\text{cut}) 
    = \delta_\text{cut}
\end{equation}
for some cutoff time $t_\text{cut}$, as shown in Fig.~\ref{scheme}(d). We evolve the master equation [Eq.~\eqref{MasterEq}] using the method of quantum Monte Carlo wave functions~\cite{Molmer}.

\begin{figure}
\centerline{\includegraphics[width=\linewidth]{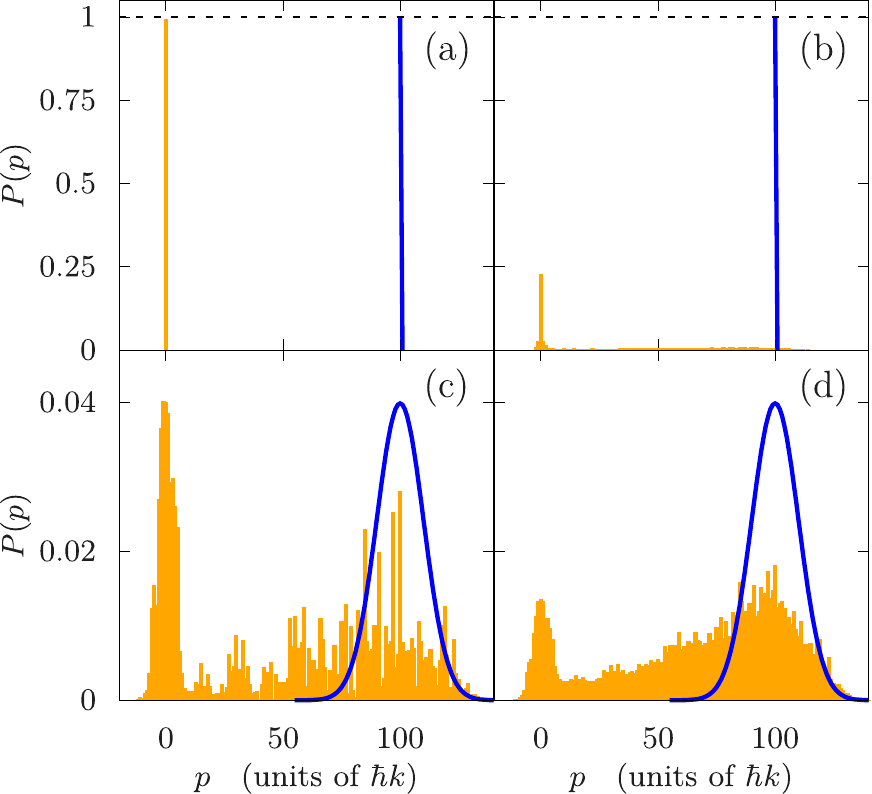}}
\caption{Initial (blue) and final (orange) momentum distributions $P(p)$ of a system subject to the slowing protocol. The system is initialized in the internal ground state $\ket{g}$ and with a momentum distribution of either the $100 \hbar k$ eigenstate (top row) or a Gaussian state with average momentum $\langle \hat p \rangle = 100 \hbar k$ and width $\sigma_p = 10 \hbar k$ (bottom row). The excited state linewidth is $\Gamma=0$ in (a) and (c), and $\Gamma = \omega_r$ in (b) and (d). Further details are discussed in text. Other parameters are: $\Omega = 100 \omega_r, \, T=0.032/\omega_r, \, t_\text{tot} = 1.6/\omega_r$, and  $\, \delta_\text{cut}=244\omega_r, \, \beta = 0.85 \, \pi /2,$ and $\bar p_0=100\hbar k$. Subplots (b) and (d) are averaged over 1,000 trajectories.}
\label{snapshots}
\end{figure}

Fig.~\ref{snapshots} presents the momentum distribution of an ensemble of particles before (blue) and after (orange) application of our slowing scheme over 100 sweeps in the case of purely coherent (left column) and dissipative (right column) dynamics. 
In the dissipative case, we chose $\Gamma = \omega_r$ as an order-of-magnitude estimate for a narrow-linewidth transition. We chose to initialize the momentum distribution as either a momentum eigenstate at $100 \hbar k$ (top row) or a Gaussian profile with an average momentum $\langle \hat p \rangle =100 \hbar k$ and a standard deviation of $\sigma_{p,0} = 10 \hbar k$ (bottom row), which typically corresponds to an initial average particle speed on the order of 1 m/s. While actual physical systems may have much higher initial particle beam speeds, this choice sufficiently demonstrates the slowing effects of our protocol. We have chosen to use the average momentum of the Gaussian as the initial momentum used in the laser frequencies ($\bar p_0=\langle \hat p \rangle$) because it results in the largest fraction of slowed particles. For experimental accessibility and robustness, we chose $\beta = 0.85 \, \pi/2$ and $\Omega = 100 \Gamma$, which results in a shortcut time $T = 0.032/\Gamma$ [see Eq.~\eqref{eq:Omega}] typically on the order of microseconds, much shorter than the associated timescale of radiation pressure slowing on the transition. This choice of $\Omega$ also sets the expected number of scattered photons $N$ to be on the order of unity [see Eq.~\eqref{numberSE}].

%p{1.25cm}||p{2.7cm}|p{2.7cm}|p{2.4cm}|p{3.4cm}|p{4.4cm}
\begin{table*}
\bgroup
\def\arraystretch{1.5}
\begin{tabular}{c||c|c|c|c|c}
 \multicolumn{6}{c}{} \\
 \hline
 Species & Transition & Wavelength (nm) & Linewidth (kHz) & Recoil frequency (kHz) & Saturation intensity (W/cm$^2$) \\
 \hline
 ${}^{40}$Ca & ${}^1S_0 \rightarrow {}^3P_1$ & 657 & 0.4 & 11.5  & $1.8 \times 10^{-7}$\\
 ${}^{88}$Sr & ${}^1S_0 \rightarrow {}^3P_1$ & 689 & 7.5 & 4.8 & $3.0 \times 10^{-6}$  \\
  YO & $X^2 \Sigma^+ \rightarrow A'^2 \Delta_{3/2}$  & 690 & 5.9 & 4.0 & $2.3 \times 10^{-6}$ \\
 ${}^{174}$Yb &  ${}^1S_0 \rightarrow {}^3P_1$ & 556 & 180 & 3.7 & $1.4 \times 10^{-4}$  \\
 BaH & $X^2 \Sigma^+ \rightarrow A^2 \Pi$ & 1061 & 1200 & 1.3 & $1.3 \times 10^{-4}$ \\
 \hline
\end{tabular}
 \caption{Fundamental properties for various atomic and molecular candidates for shortcut slowing on a narrow transition.}
 \label{candidateTable1}
 \egroup
\end{table*}

\begin{table*}
\bgroup
\def\arraystretch{1.5}
\begin{tabular}{c||c|c|c|c|c}
 \multicolumn{6}{c}{} \\
 \hline
 Species & Rabi frequency (MHz) & Initial speed at 200$\hbar k$ (m/s) & Slowing time ($\mu$s) & Slowing distance ($\mu$m) & Capture fraction \\
 \hline
 ${}^{40}$Ca  & 0.2 &  3.0 & 494 & 746 & $ 5.6 \times 10^{-1}$  \\
 ${}^{88}$Sr  & 1.0 & 1.3 & 106 & 70 & $ 1.5 \times 10^{-1}$ \\
 YO & 0.9 & 1.1 & 119 & 66 & $2.1 \times 10^{-1} $ \\
 ${}^{174}$Yb & 3.4 & 0.82 & 30 & 12 & $6.8\times 10^{-6}$ \\
 BaH & 23.6 & 0.54 & 4 & 1 & $1.6 \times 10^{-5}$\\
 \hline
 \end{tabular}
 \egroup
 \caption{Slowing results for various atomic and molecular candidates using a 0.1 W/cm$^2$ laser intensity and starting from a momentum of $200 \hbar k$. The Rabi frequency, slowing time, and slowing distance follow from Eqns.~\eqref{rabiIntensity}, $\eqref{slowingTime}$, and~\eqref{xLRIOmega} respectively, with $\beta = 0.85 \, \pi /2$. The simulated capture fraction, defined as the fraction of particles with momentum $|p| \leq 3 \hbar k$ after the slowing process, was calculated over 1,000 trajectories for Ca, Sr, and YO, and 10,000 trajectories for Yb and BaH.}
 \label{candidateTable2}
 \end{table*}

As seen in Fig.~\ref{snapshots}, a substantial fraction of the distribution is slowed to near zero momentum. We first discuss the results in the case of purely coherent dynamics (left column). If the system begins in the eigenstate $\ket{g,100\hbar k}$ [Fig.~\ref{snapshots}(a)], about 99.3$\%$ of the population ends in the zero momentum eigenstate. (We attribute the lack of 100$\%$ transfer to our use of a cutoff frequency $\delta_\text{cut} = 244 \omega_r$.) If the system begins in a Gaussian state [Fig.~\ref{snapshots}(c)], about 35$\%$ of the population ends with momentum $|p| \leq 10 \hbar k$, which corresponds to about half of the population within one standard deviation of the average momentum in the initial distribution. 

In the case of dissipative dynamics, 29$\%$ of the population ends with momentum $|p| \leq \hbar k$ if the system is initialized in the eigenstate $\ket{g,100\hbar k}$ [Fig.~\ref{snapshots}(b)], and 13$\%$ of the population ends with momentum $|p| \leq 10 \hbar k$ if the system is initialized in the Gaussian state [Fig.~\ref{snapshots}(d)]. In both cases, there was an average of 1.6 scattered photons per particle, which agrees with the predicted scattering rate given in Table~\ref{dynamicsTable}.
These results demonstrate that our protocol can potentially slow a significant fraction of particles to zero momentum in a very short distance and with few spontaneous emission events. An appealing feature of our protocol is its ability to apply high slowing forces with a virtually negligible reliance on spontaneous emission.

As a more realistic demonstration, we present several atomic and molecular species that are reasonable candidates for our slowing protocol in Table~\ref{candidateTable1}. We simulated the slowing of each species with an initial momentum of $200 \hbar k$ using a 0.1 W/cm$^2$ laser intensity, and provide the expected slowing times $\Delta t$, slowing distances $\Delta x$, and capture fractions $C$ in Table~\ref{candidateTable2}. The saturation intensity $I_\text{sat}$, laser intensity $I$, and Rabi frequency $\Omega$ can be calculated according to \cite{Metcalf}
\begin{equation}
\label{rabiIntensity}
 2 \left( \frac{\Omega}{\Gamma} \right)^2 = \frac{I}{I_\text{sat}}; 
 \quad I_\text{sat} \equiv \frac{\pi h c \Gamma }{3 \lambda^3},
\end{equation}
where $h$ is Planck's constant and $c$ is the speed of light. 
The calculated slowing times and distances become smaller as the transition dipole matrix element increases, as can be seen for the species with larger linewidths. However, the required temporal control of the Rabi frequency and detuning profiles in these cases may be an experimental challenge. One solution is to simply reduce the laser power, but this can quickly move the parameters away from the regime $\Omega \gg \Gamma$, which increases the chance of spontaneous emission and hence reduces the capture fraction $C$.

Due to our method's sensitivity to spontaneous emission, a rough estimate for the capture fraction $C$ is given by the fraction of particles that do not emit a single spontaneous photon throughout the slowing process. Using Eq.~\eqref{numberSE}, this is approximately
\begin{equation}
\label{captureFrac}
    C \approx 
    \exp\left(-N_\text{LRI}\right)
    =
    \exp \left(
    -\frac{\Gamma}{\Omega} \frac{\pi}{2 \sin \beta} \zeta_0
    \right).
\end{equation}
This formula is on the same order as the simulated capture fractions given in Table~\ref{candidateTable2} for Ca, Sr, and YO, but it underestimates the results for Yb and BaH, suggesting that particles can still be slowed after emitting spontaneous photons. Eq.~\eqref{captureFrac} is not a fundamental limit, as it can be improved with a more sophisticated implementation of the shortcut process, such as introducing occasional waiting periods so that the particles return to the internal ground state (as is done in other methods \cite{ARP1}), allowing the laser pulses to overlap in time (see Fig.~\ref{overlap}), or repeating the slowing protocol over a range of momentum states. For example, we were able to increase the simulated ${}^{174}$Yb capture fraction by over an order of magnitude to $C=1.7 \times 10^{-4}$ by introducing a laser pulse overlap fraction $f=0.2$ as discussed in Section~\ref{Robustness}.

\begin{figure}
\centerline{\includegraphics[width=\linewidth]{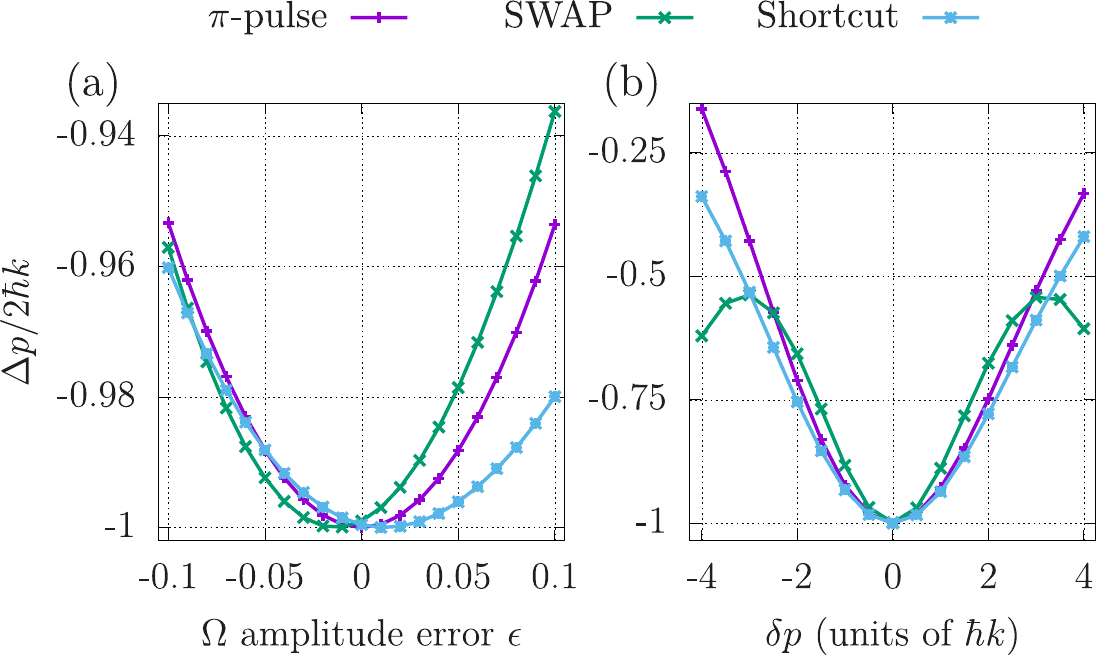}}
\caption{Robustness comparison of $\pi$-pulse slowing (purple, plus), SWAP slowing (green, cross), and the shortcut scheme (blue, circle with cross) over a $\bar p \rightarrow \bar p - 2 \hbar k$ transfer process. Ideally, the resulting impulse $\Delta p$ satisfies $\Delta p/ 2 \hbar k = -1$. The evolution is purely coherent ($\Gamma =0$), the momentum of the particle (which is addressed by the laser frequencies) is $\bar p = 2 \hbar k$, and the Rabi frequency is $\Omega = 10 \omega_r$ for all processes. (a) Impulse $\Delta p$ experienced by the particle, in units of the ideal impulse magnitude $2 \hbar k$, as a function of the error in the Rabi frequency amplitude $\epsilon$ [see Eq.~\eqref{rabiError}]. The shortcut scheme is the most robust protocol when $\epsilon > 0$. (b) $\Delta p/ 2 \hbar k$ as a function of the relative momentum of the particle $\delta p$ with respect to the momentum $\bar p = 2 \hbar k$ used in the laser frequencies [see Eqns.~\eqref{eq:laserFreqs} and~\eqref{relativeMom}]. For this set of parameters, the shortcut and $\pi$-pulse scheme are generally more robust than SWAP slowing. $\pi$-pulse parameters are: $T = 0.314/\omega_r$ and $\beta = \pi /2$. SWAP slowing parameters are:  $T = 1/\omega_r$ and $\Delta = 50\omega_r$. Shortcut parameters are: $T = 0.44/\omega_r, \, \delta_\text{cut} = 230\omega_r$, and $\beta = 0.5 \, \pi /2$.}
\label{robustness}
\end{figure}

\section{Robustness} \label{Robustness}

\begin{figure}
\centerline{\includegraphics[width=\linewidth]{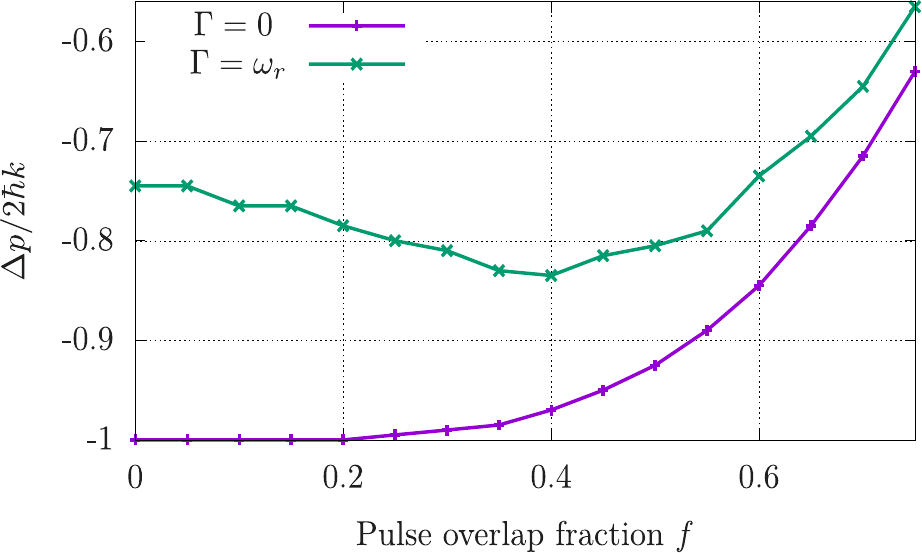}}
\caption{Impulse $\Delta p$ experienced by the particle, in units of the ideal impulse magnitude $2 \hbar k$, as a function of the laser pulse overlap fraction $f$. The impulse is calculated under both coherent ($\Gamma = 0$) and dissipative ($\Gamma = \omega_r$) dynamics. Other parameters are: $T = 0.44/\omega_r, \, \delta_\text{cut} = 230\omega_r, \, \Omega = 10\omega_r, \, \bar p = 2 \hbar k$, and $\beta = 0.5 \, \pi /2$. All points are averaged over 1,000 trajectories.}
\label{overlap}
\end{figure}

We now study the robustness of the shortcut slowing scheme to various systematic errors that may arise in an experimental setting.
Specifically, we modify the Rabi frequency amplitude, then separately consider the result of slowing a particle with a momentum $p$ that is not equal to the momentum accounted for in the laser frequencies $\bar p$ (Fig.~\ref{robustness}).
We also consider the scheme's robustness to the temporal overlap $f$ of the square pulses (Fig.~\ref{overlap}), which may minimize scattering events since it potentially reduces the amount of time the particle remains in the internal excited state.
For simplicity, we focus on a single $\ket{g} \rightarrow \ket{e} \rightarrow \ket{g}$ process and calculate the resulting impulse $\Delta p$ experienced by the particle. Moreover, we employ a phase relation between the laser pulses such that the Rabi frequency of each pulse is purely real. The correct detuning profiles are employed throughout this section. We set the Rabi frequency to $\Omega = 10 \omega_r$ across all slowing methods, but necessarily allow for different slowing times.

Using the auxiliary variable $\beta = 0.5 \, \pi/2$, we compare the robustness of our scheme to both $\pi$-pulse and SWAP slowing.
As discussed in Section~\ref{bloch}, The $\pi$-pulse solution is a special case of our slowing method with the choice $\beta = \pi/2$, which results in a fixed detuning [see Eq.~\eqref{eq:LRIdetuning}].
SWAP slowing is simulated by using a sawtooth-wave detuning profile with full range $\Delta$ and period $T$ for each single-photon transfer. Unlike SWAP cooling, the lasers are sequentially pulsed as in Fig.~\ref{scheme}(a) and~\ref{scheme}(b), and the laser detunings are centered on the momentum $\bar p$. 

While there is not a unique way to choose the SWAP slowing parameters, we used the following method. Landau and Zener showed~\cite{zener} that the population of the second (initially unoccupied) state $P_2$ if the laser is linearly chirped from a detuning of minus infinity to positive infinity, is
\begin{equation}
\label{LZprob}
    P_2 = 1 - \exp
    \left(
    - \frac{\pi}{2} \frac{\Omega^2}{\alpha} \right),
\end{equation}
where $\alpha$ is the frequency chirping rate in rad/s$^2$.
Therefore, to obtain a transfer probability of at least 95\%, we chose to set $\Omega^2/\alpha=2$. Next, we increased the chance of population transfer by setting the shortcut period to be several times larger than the approximate time $\tau_j$ required to transfer population between quantum states in the adiabatic regime~\cite{vitanov}:
\begin{equation}
  \tau_j = \frac{2 \Omega}{\alpha} = \frac{0.4}{\omega_r}
  \quad \Rightarrow \quad
  T = \frac{1}{\omega_r} > \tau_j.
\end{equation}
These choices constrained the sweep range to be $\Delta = 50 \omega_r$. Note that the resulting SWAP slowing period $T=1/\omega_r$ is significantly longer than the $\pi$-pulse ($T \approx 0.3/\omega_r$) and shortcut slowing ($T \approx 0.4/\omega_r$) periods [see Eq.~\eqref{eq:Omega}].

Fig.~\ref{robustness}(a) presents the resulting particle impulse $\Delta p$ as a function of the error in the Rabi frequency amplitude, which is characterized by the small parameter~$\epsilon$:
\begin{equation}
\label{rabiError}
    \Omega \rightarrow \Omega(1+\epsilon).
\end{equation}
While the shortcut scheme applies a similar impulse $\Delta p$ compared to the $\pi$-pulse method for $\epsilon <0$, it is the most robust method for $\epsilon>0$. 
The SWAP slowing result, while being the most robust method for $\epsilon < 0$, can change significantly with small changes to the parameters. This instability is as a result of high-frequency population oscillations \cite{vitanov}. Moreover, when dissipation is included, the much longer SWAP slowing period increases the chance of spontaneous emission, which can disrupt the slowing process. These results demonstrate the utility of the shortcut scheme, as it is robust to small errors $\epsilon$ and takes much less time than SWAP slowing. Note that intensity modulators typically have errors less than approximately 5\%. 

Fig.~\ref{robustness}(b) presents the effects of applying each slowing method to a particle with a momentum $p$ which is not equal to the momentum $\bar p$ accounted for in the laser frequencies [see Eq.~\eqref{eq:laserFreqs}]. Such an error occurs when slowing a cloud of particles with a distribution of momenta. We parameterize this difference in momentum with the variable $\delta p$ as:
\begin{equation}
    \label{relativeMom}
    p = \bar p + \delta p.
\end{equation}
In this case, we find that the shortcut scheme is most robust for small $\delta p$, but SWAP slowing becomes more robust for $|\delta p| > 3 \hbar k$. This change in the trend for SWAP slowing near the particular values $\delta p = \pm 3 \hbar k$ is due to population oscillations, and the results can again change significantly with small changes in the parameters. In the limit of adiabatic dynamics and time-resolved transfers (which necessarily takes a long time), SWAP slowing is generally most robust to this error since the rate of change of the detuning profile in SWAP slowing is constant, thereby removing the need to align the center of the laser detuning profile with the Doppler shift of the particle. It should be noted that it is not necessarily a problem that the shortcut scheme is less robust to large deviations of $\delta p$ from zero because we scan through values of $p/ \hbar k$ in integer steps by design. What is more important is that the shortcut scheme is more robust when $|\delta p| < \hbar k$ than SWAP slowing, as particles will not necessarily have integer momentum.

Fig.~\ref{overlap} presents the effects of allowing the laser pulses to overlap in time. We define the overlap fraction of the square pulses $f$ such that $f=0$ when the pulses are completely time-resolved but occur sequentially with no delay, and $f=1$ when the pulses occur at the same time for the entire pulse duration.  We find, in the case of purely coherent dynamics ($\Gamma=0$), that a pulse overlap fraction as large as $f=0.2$ does not change the impulse experienced by the particle. In the case of dissipative dynamics, we find that there is an optimal $f$ that maximizes the impulse $\Delta p$. This optimal $f$, which depends on system specifics, must be small enough to allow transfer to $\ket{e}$ before the transfer back to $\ket{g}$, but large enough to minimize the time the particle spends in the excited state, thereby minimizing the chance of spontaneous emission. It should be noted that a pulse overlap introduces the possibility of multiphoton, or Doppleron, resonances \cite{dopplerons} which may interfere with the single-photon slowing dynamics.

When slowing a particle with a large initial momentum $p \gg \hbar k$ to rest, we emphasize that the slowing efficiency is significantly affected by even a small deviation from the ideal impulse $\Delta p = - 2 \hbar k$ since the error compounds exponentially with the number of transfer processes, and the particle is generally not transferred back to $\ket{g}$ for the next pulse sequence. It may be possible to further enhance the robustness of the slowing protocol to the rapidly changing detuning profile at the beginning and end of the transfers by using a Rabi frequency profile that satisfies $\Omega(t_0) = \Omega(t_f) \rightarrow 0$, such as a Gaussian or a sinusoidal function \cite{eberly,ARP1,ARP2}. However, in order to satisfy the auxiliary equations and boundary conditions [see Eqns.~\eqref{AuxEq1}-\eqref{eq:gammaBoundary}], the peak Rabi frequency may need to be larger than what we consider here.

\section{Conclusion and Outlook}
In this work, we proposed a purely coherent particle slowing scheme. By utilizing the method of inverse-engineering based on Lewis-Riesenfeld invariant theory, we demonstrated that these slowing designs are able to achieve effective adiabatic dynamics but with a short slowing time. We illustrated how to design Rabi frequency and detuning profiles that achieve the desired dynamics and demonstrated that our slowing blueprint is a promising alternative to conventional slowing schemes that rely on radiation pressure for narrow linewidth systems or systems that lack a closed cycling transition. We theoretically examined the effective classical forces exerted on the particles during the slowing processes and demonstrated that, when the Rabi frequency is large compared to the excited state decay rate, our scheme is able to exert significantly higher forces than radiation pressure while maintaining a very low number of scattered photons.
We also showed that our proposed slowing scheme is robust to various systematic errors.

A possible practical implementation of our scheme is to directly apply it to a particle beam exiting a supersonic nozzle or buffer gas cell. Our scheme may also be utilized as a second slowing stage after particles exiting an effusive oven have been slowed in a precursor stage to the order of $10 \text{ m/s}$ and the initial spread in velocity has been greatly reduced. Additionally, it may be possible to implement our protocol as a steady-state slowing procedure by compensating for the changing particle velocity with a magnetic field gradient instead of the time-dependent laser frequencies, in a similar approach to a Stark or Zeeman decelerator.

A natural next step would be to optimize our shortcut solutions with respect to different criteria. 
Time-optimization of different adiabatic shortcuts has been studied previously for two-level systems \cite{Control}, STIRAP \cite{Mortensen}, and frictionless cooling in harmonic traps \cite{Stefanatos}. Another optimization criterion comes from the fact that adiabatic shortcuts cannot be implemented without an energetic cost, an intrinsic relationship that has been rigorously studied for Berry's transitionless quantum driving algorithm \cite{Campbell}, various other shortcuts \cite{Abah}, and in its applications to quantum computing \cite{Coulamy}. Thus, minimizing the energetic cost for a fixed sweep period is, in effect, finding the most efficient shortcut.

Another appealing scheme is a similar shortcut protocol applied to a system with an internal state structure comprised of two stable ground states coupled to an excited state, such as found in a $J=1 \rightarrow J'=1$ transition, where $J$ is the total electronic angular momentum quantum number.  Speeding up a Raman transition between the ground states could remove twice as much momentum per transition. This model, while generally requiring more laser power, has the additional benefit of an engineered excited state linewidth, which can be made arbitrarily small if the lasers are sufficiently detuned from the excited state. Moreover, applying anti-symmetric detuning sweeps could allow for control of the slowed velocity range, potentially resulting in a slowed distribution with a low temperature.

There are similarities between our results and others \cite{eberly,MetcalfNonAdiabatic,ARP1,ARP2} in the sense that two-level inversion is achieved in the diabatic limit. A more thorough investigation could further our understanding of the connection between these solutions and adiabatic shortcuts.

Furthermore, it is impossible to implement the LRI scheme to the entire momentum Hilbert space by hand. An intriguing possible solution to this problem is to employ advanced optimization techniques, such as reinforcement learning, to find a shortcut solution that maximizes the slowing scheme's capture range or minimizes the energetic cost of implementing the shortcut for a given sweep period. 

\section{Acknowledgments}
We would like to thank Athreya Shankar, John Cooper, Matt Norcia, and Shiqian Ding for useful discussions. This work was supported by NSF PFC Grant No. PHY 1734006 and NSF Grant No. PHY 1806827.

\bibliography{references.bib}

%merlin.mbs apsrev4-1.bst 2010-07-25 4.21a (PWD, AO, DPC) hacked
%Control: key (0)
%Control: author (72) initials jnrlst
%Control: editor formatted (1) identically to author
%Control: production of article title (-1) disabled
%Control: page (0) single
%Control: year (1) truncated
%Control: production of eprint (0) enabled
\begin{thebibliography}{59}%
\makeatletter
\providecommand \@ifxundefined [1]{%
 \@ifx{#1\undefined}
}%
\providecommand \@ifnum [1]{%
 \ifnum #1\expandafter \@firstoftwo
 \else \expandafter \@secondoftwo
 \fi
}%
\providecommand \@ifx [1]{%
 \ifx #1\expandafter \@firstoftwo
 \else \expandafter \@secondoftwo
 \fi
}%
\providecommand \natexlab [1]{#1}%
\providecommand \enquote  [1]{``#1''}%
\providecommand \bibnamefont  [1]{#1}%
\providecommand \bibfnamefont [1]{#1}%
\providecommand \citenamefont [1]{#1}%
\providecommand \href@noop [0]{\@secondoftwo}%
\providecommand \href [0]{\begingroup \@sanitize@url \@href}%
\providecommand \@href[1]{\@@startlink{#1}\@@href}%
\providecommand \@@href[1]{\endgroup#1\@@endlink}%
\providecommand \@sanitize@url [0]{\catcode `\\12\catcode `\$12\catcode
  `\&12\catcode `\#12\catcode `\^12\catcode `\_12\catcode `\%12\relax}%
\providecommand \@@startlink[1]{}%
\providecommand \@@endlink[0]{}%
\providecommand \url  [0]{\begingroup\@sanitize@url \@url }%
\providecommand \@url [1]{\endgroup\@href {#1}{\urlprefix }}%
\providecommand \urlprefix  [0]{URL }%
\providecommand \Eprint [0]{\href }%
\providecommand \doibase [0]{http://dx.doi.org/}%
\providecommand \selectlanguage [0]{\@gobble}%
\providecommand \bibinfo  [0]{\@secondoftwo}%
\providecommand \bibfield  [0]{\@secondoftwo}%
\providecommand \translation [1]{[#1]}%
\providecommand \BibitemOpen [0]{}%
\providecommand \bibitemStop [0]{}%
\providecommand \bibitemNoStop [0]{.\EOS\space}%
\providecommand \EOS [0]{\spacefactor3000\relax}%
\providecommand \BibitemShut  [1]{\csname bibitem#1\endcsname}%
\let\auto@bib@innerbib\@empty
%</preamble>
\bibitem [{\citenamefont {Chu}(1998)}]{chu}%
  \BibitemOpen
  \bibfield  {author} {\bibinfo {author} {\bibfnamefont {S.}~\bibnamefont
  {Chu}},\ }\href {\doibase 10.1103/RevModPhys.70.685} {\bibfield  {journal}
  {\bibinfo  {journal} {Rev. Mod. Phys.}\ }\textbf {\bibinfo {volume} {70}},\
  \bibinfo {pages} {685} (\bibinfo {year} {1998})}\BibitemShut {NoStop}%
\bibitem [{\citenamefont {Phillips}(1998)}]{phillips}%
  \BibitemOpen
  \bibfield  {author} {\bibinfo {author} {\bibfnamefont {W.~D.}\ \bibnamefont
  {Phillips}},\ }\href {\doibase 10.1103/RevModPhys.70.721} {\bibfield
  {journal} {\bibinfo  {journal} {Rev. Mod. Phys.}\ }\textbf {\bibinfo {volume}
  {70}},\ \bibinfo {pages} {721} (\bibinfo {year} {1998})}\BibitemShut
  {NoStop}%
\bibitem [{\citenamefont {Cohen-Tannoudji}(1998)}]{CT}%
  \BibitemOpen
  \bibfield  {author} {\bibinfo {author} {\bibfnamefont {C.~N.}\ \bibnamefont
  {Cohen-Tannoudji}},\ }\href {\doibase 10.1103/RevModPhys.70.707} {\bibfield
  {journal} {\bibinfo  {journal} {Rev. Mod. Phys.}\ }\textbf {\bibinfo {volume}
  {70}},\ \bibinfo {pages} {707} (\bibinfo {year} {1998})}\BibitemShut
  {NoStop}%
\bibitem [{\citenamefont {Metcalf}\ and\ \citenamefont
  {Straten}(2002)}]{Metcalf}%
  \BibitemOpen
  \bibfield  {author} {\bibinfo {author} {\bibfnamefont {H.~J.}\ \bibnamefont
  {Metcalf}}\ and\ \bibinfo {author} {\bibfnamefont {P.~v.~d.}\ \bibnamefont
  {Straten}},\ }\href@noop {} {\emph {\bibinfo {title} {Laser cooling and
  trapping}}}\ (\bibinfo  {publisher} {Springer},\ \bibinfo {year}
  {2002})\BibitemShut {NoStop}%
\bibitem [{\citenamefont {Safronova}\ \emph {et~al.}(2018)\citenamefont
  {Safronova}, \citenamefont {Budker}, \citenamefont {DeMille}, \citenamefont
  {Kimball}, \citenamefont {Derevianko},\ and\ \citenamefont
  {Clark}}]{Safronova}%
  \BibitemOpen
  \bibfield  {author} {\bibinfo {author} {\bibfnamefont {M.~S.}\ \bibnamefont
  {Safronova}}, \bibinfo {author} {\bibfnamefont {D.}~\bibnamefont {Budker}},
  \bibinfo {author} {\bibfnamefont {D.}~\bibnamefont {DeMille}}, \bibinfo
  {author} {\bibfnamefont {D.~F.~J.}\ \bibnamefont {Kimball}}, \bibinfo
  {author} {\bibfnamefont {A.}~\bibnamefont {Derevianko}}, \ and\ \bibinfo
  {author} {\bibfnamefont {C.~W.}\ \bibnamefont {Clark}},\ }\href {\doibase
  10.1103/RevModPhys.90.025008} {\bibfield  {journal} {\bibinfo  {journal}
  {Rev. Mod. Phys.}\ }\textbf {\bibinfo {volume} {90}},\ \bibinfo {pages}
  {025008} (\bibinfo {year} {2018})}\BibitemShut {NoStop}%
\bibitem [{\citenamefont {Anderson}\ \emph {et~al.}(1995)\citenamefont
  {Anderson}, \citenamefont {Ensher}, \citenamefont {Matthews}, \citenamefont
  {Wieman},\ and\ \citenamefont {Cornell}}]{Cornell}%
  \BibitemOpen
  \bibfield  {author} {\bibinfo {author} {\bibfnamefont {M.~H.}\ \bibnamefont
  {Anderson}}, \bibinfo {author} {\bibfnamefont {J.~R.}\ \bibnamefont
  {Ensher}}, \bibinfo {author} {\bibfnamefont {M.~R.}\ \bibnamefont
  {Matthews}}, \bibinfo {author} {\bibfnamefont {C.~E.}\ \bibnamefont
  {Wieman}}, \ and\ \bibinfo {author} {\bibfnamefont {E.~A.}\ \bibnamefont
  {Cornell}},\ }\href {\doibase 10.1126/science.269.5221.198} {\bibfield
  {journal} {\bibinfo  {journal} {Science}\ }\textbf {\bibinfo {volume}
  {269}},\ \bibinfo {pages} {198} (\bibinfo {year} {1995})},\ \Eprint
  {http://arxiv.org/abs/https://science.sciencemag.org/content/269/5221/198.\\full.pdf}
  {https://science.sciencemag.org/content/269/5221/198.\\full.pdf} \BibitemShut
  {NoStop}%
\bibitem [{\citenamefont {Regal}\ \emph {et~al.}(2004)\citenamefont {Regal},
  \citenamefont {Greiner},\ and\ \citenamefont {Jin}}]{Jin}%
  \BibitemOpen
  \bibfield  {author} {\bibinfo {author} {\bibfnamefont {C.~A.}\ \bibnamefont
  {Regal}}, \bibinfo {author} {\bibfnamefont {M.}~\bibnamefont {Greiner}}, \
  and\ \bibinfo {author} {\bibfnamefont {D.~S.}\ \bibnamefont {Jin}},\ }\href
  {\doibase 10.1103/PhysRevLett.92.040403} {\bibfield  {journal} {\bibinfo
  {journal} {Phys. Rev. Lett.}\ }\textbf {\bibinfo {volume} {92}},\ \bibinfo
  {pages} {040403} (\bibinfo {year} {2004})}\BibitemShut {NoStop}%
\bibitem [{\citenamefont {Kim}\ \emph {et~al.}(2010)\citenamefont {Kim},
  \citenamefont {Chang}, \citenamefont {Korenblit}, \citenamefont {Islam},
  \citenamefont {Edwards}, \citenamefont {Freericks}, \citenamefont {Lin},
  \citenamefont {Duan},\ and\ \citenamefont {Monroe}}]{Kim_QS}%
  \BibitemOpen
  \bibfield  {author} {\bibinfo {author} {\bibfnamefont {K.}~\bibnamefont
  {Kim}}, \bibinfo {author} {\bibfnamefont {M.-S.}\ \bibnamefont {Chang}},
  \bibinfo {author} {\bibfnamefont {S.}~\bibnamefont {Korenblit}}, \bibinfo
  {author} {\bibfnamefont {R.}~\bibnamefont {Islam}}, \bibinfo {author}
  {\bibfnamefont {E.~E.}\ \bibnamefont {Edwards}}, \bibinfo {author}
  {\bibfnamefont {J.~K.}\ \bibnamefont {Freericks}}, \bibinfo {author}
  {\bibfnamefont {G.-D.}\ \bibnamefont {Lin}}, \bibinfo {author} {\bibfnamefont
  {L.-M.}\ \bibnamefont {Duan}}, \ and\ \bibinfo {author} {\bibfnamefont
  {C.}~\bibnamefont {Monroe}},\ }\href {\doibase 10.1038/nature09071}
  {\bibfield  {journal} {\bibinfo  {journal} {Nature}\ }\textbf {\bibinfo
  {volume} {465}},\ \bibinfo {pages} {590–593} (\bibinfo {year}
  {2010})}\BibitemShut {NoStop}%
\bibitem [{\citenamefont {Narevicius}\ \emph {et~al.}(2008)\citenamefont
  {Narevicius}, \citenamefont {Libson}, \citenamefont {Parthey}, \citenamefont
  {Chavez}, \citenamefont {Narevicius}, \citenamefont {Even},\ and\
  \citenamefont {Raizen}}]{Narevicius}%
  \BibitemOpen
  \bibfield  {author} {\bibinfo {author} {\bibfnamefont {E.}~\bibnamefont
  {Narevicius}}, \bibinfo {author} {\bibfnamefont {A.}~\bibnamefont {Libson}},
  \bibinfo {author} {\bibfnamefont {C.~G.}\ \bibnamefont {Parthey}}, \bibinfo
  {author} {\bibfnamefont {I.}~\bibnamefont {Chavez}}, \bibinfo {author}
  {\bibfnamefont {J.}~\bibnamefont {Narevicius}}, \bibinfo {author}
  {\bibfnamefont {U.}~\bibnamefont {Even}}, \ and\ \bibinfo {author}
  {\bibfnamefont {M.~G.}\ \bibnamefont {Raizen}},\ }\href {\doibase
  10.1103/PhysRevLett.100.093003} {\bibfield  {journal} {\bibinfo  {journal}
  {Phys. Rev. Lett.}\ }\textbf {\bibinfo {volume} {100}},\ \bibinfo {pages}
  {093003} (\bibinfo {year} {2008})}\BibitemShut {NoStop}%
\bibitem [{\citenamefont {Akerman}\ \emph {et~al.}(2015)\citenamefont
  {Akerman}, \citenamefont {Karpov}, \citenamefont {David}, \citenamefont
  {Lavert-Ofir}, \citenamefont {Narevicius},\ and\ \citenamefont
  {Narevicius}}]{Akerman}%
  \BibitemOpen
  \bibfield  {author} {\bibinfo {author} {\bibfnamefont {N.}~\bibnamefont
  {Akerman}}, \bibinfo {author} {\bibfnamefont {M.}~\bibnamefont {Karpov}},
  \bibinfo {author} {\bibfnamefont {L.}~\bibnamefont {David}}, \bibinfo
  {author} {\bibfnamefont {E.}~\bibnamefont {Lavert-Ofir}}, \bibinfo {author}
  {\bibfnamefont {J.}~\bibnamefont {Narevicius}}, \ and\ \bibinfo {author}
  {\bibfnamefont {E.}~\bibnamefont {Narevicius}},\ }\href {\doibase
  10.1088/1367-2630/17/6/065015} {\bibfield  {journal} {\bibinfo  {journal}
  {New Journal of Physics}\ }\textbf {\bibinfo {volume} {17}},\ \bibinfo
  {pages} {065015} (\bibinfo {year} {2015})}\BibitemShut {NoStop}%
\bibitem [{\citenamefont {Hogan}\ \emph {et~al.}(2008)\citenamefont {Hogan},
  \citenamefont {Wiederkehr}, \citenamefont {Schmutz},\ and\ \citenamefont
  {Merkt}}]{Hogan}%
  \BibitemOpen
  \bibfield  {author} {\bibinfo {author} {\bibfnamefont {S.~D.}\ \bibnamefont
  {Hogan}}, \bibinfo {author} {\bibfnamefont {A.~W.}\ \bibnamefont
  {Wiederkehr}}, \bibinfo {author} {\bibfnamefont {H.}~\bibnamefont {Schmutz}},
  \ and\ \bibinfo {author} {\bibfnamefont {F.}~\bibnamefont {Merkt}},\ }\href
  {\doibase 10.1103/PhysRevLett.101.143001} {\bibfield  {journal} {\bibinfo
  {journal} {Phys. Rev. Lett.}\ }\textbf {\bibinfo {volume} {101}},\ \bibinfo
  {pages} {143001} (\bibinfo {year} {2008})}\BibitemShut {NoStop}%
\bibitem [{\citenamefont {Petzold}\ \emph {et~al.}(2018)\citenamefont
  {Petzold}, \citenamefont {Kaebert}, \citenamefont {Gersema}, \citenamefont
  {Siercke},\ and\ \citenamefont {Ospelkaus}}]{Petzold}%
  \BibitemOpen
  \bibfield  {author} {\bibinfo {author} {\bibfnamefont {M.}~\bibnamefont
  {Petzold}}, \bibinfo {author} {\bibfnamefont {P.}~\bibnamefont {Kaebert}},
  \bibinfo {author} {\bibfnamefont {P.}~\bibnamefont {Gersema}}, \bibinfo
  {author} {\bibfnamefont {M.}~\bibnamefont {Siercke}}, \ and\ \bibinfo
  {author} {\bibfnamefont {S.}~\bibnamefont {Ospelkaus}},\ }\href {\doibase
  10.1088/1367-2630/aab9f5} {\bibfield  {journal} {\bibinfo  {journal} {New
  Journal of Physics}\ }\textbf {\bibinfo {volume} {20}},\ \bibinfo {pages}
  {042001} (\bibinfo {year} {2018})}\BibitemShut {NoStop}%
\bibitem [{\citenamefont {Chervenkov}\ \emph {et~al.}(2014)\citenamefont
  {Chervenkov}, \citenamefont {Wu}, \citenamefont {Bayerl}, \citenamefont
  {Rohlfes}, \citenamefont {Gantner}, \citenamefont {Zeppenfeld},\ and\
  \citenamefont {Rempe}}]{Chervenkov}%
  \BibitemOpen
  \bibfield  {author} {\bibinfo {author} {\bibfnamefont {S.}~\bibnamefont
  {Chervenkov}}, \bibinfo {author} {\bibfnamefont {X.}~\bibnamefont {Wu}},
  \bibinfo {author} {\bibfnamefont {J.}~\bibnamefont {Bayerl}}, \bibinfo
  {author} {\bibfnamefont {A.}~\bibnamefont {Rohlfes}}, \bibinfo {author}
  {\bibfnamefont {T.}~\bibnamefont {Gantner}}, \bibinfo {author} {\bibfnamefont
  {M.}~\bibnamefont {Zeppenfeld}}, \ and\ \bibinfo {author} {\bibfnamefont
  {G.}~\bibnamefont {Rempe}},\ }\href {\doibase 10.1103/PhysRevLett.112.013001}
  {\bibfield  {journal} {\bibinfo  {journal} {Phys. Rev. Lett.}\ }\textbf
  {\bibinfo {volume} {112}},\ \bibinfo {pages} {013001} (\bibinfo {year}
  {2014})}\BibitemShut {NoStop}%
\bibitem [{\citenamefont {Bethlem}\ \emph {et~al.}(2000)\citenamefont
  {Bethlem}, \citenamefont {Berden}, \citenamefont {Crompvoets}, \citenamefont
  {Jongma}, \citenamefont {van Roij},\ and\ \citenamefont {Meijer}}]{Bethlem}%
  \BibitemOpen
  \bibfield  {author} {\bibinfo {author} {\bibfnamefont {H.~L.}\ \bibnamefont
  {Bethlem}}, \bibinfo {author} {\bibfnamefont {G.}~\bibnamefont {Berden}},
  \bibinfo {author} {\bibfnamefont {F.~M.~H.}\ \bibnamefont {Crompvoets}},
  \bibinfo {author} {\bibfnamefont {R.~T.}\ \bibnamefont {Jongma}}, \bibinfo
  {author} {\bibfnamefont {A.~J.~A.}\ \bibnamefont {van Roij}}, \ and\ \bibinfo
  {author} {\bibfnamefont {G.}~\bibnamefont {Meijer}},\ }\href {\doibase
  10.1038/35020030} {\bibfield  {journal} {\bibinfo  {journal} {Nature}\
  }\textbf {\bibinfo {volume} {406}},\ \bibinfo {pages} {491} (\bibinfo {year}
  {2000})}\BibitemShut {NoStop}%
\bibitem [{\citenamefont {Truppe}\ \emph {et~al.}(2017)\citenamefont {Truppe},
  \citenamefont {Williams}, \citenamefont {Fitch}, \citenamefont {Hambach},
  \citenamefont {Wall}, \citenamefont {Hinds}, \citenamefont {Sauer},\ and\
  \citenamefont {Tarbutt}}]{Truppe}%
  \BibitemOpen
  \bibfield  {author} {\bibinfo {author} {\bibfnamefont {S.}~\bibnamefont
  {Truppe}}, \bibinfo {author} {\bibfnamefont {H.~J.}\ \bibnamefont
  {Williams}}, \bibinfo {author} {\bibfnamefont {N.~J.}\ \bibnamefont {Fitch}},
  \bibinfo {author} {\bibfnamefont {M.}~\bibnamefont {Hambach}}, \bibinfo
  {author} {\bibfnamefont {T.~E.}\ \bibnamefont {Wall}}, \bibinfo {author}
  {\bibfnamefont {E.~A.}\ \bibnamefont {Hinds}}, \bibinfo {author}
  {\bibfnamefont {B.~E.}\ \bibnamefont {Sauer}}, \ and\ \bibinfo {author}
  {\bibfnamefont {M.~R.}\ \bibnamefont {Tarbutt}},\ }\href {\doibase
  10.1088/1367-2630/aa5ca2} {\bibfield  {journal} {\bibinfo  {journal} {New
  Journal of Physics}\ }\textbf {\bibinfo {volume} {19}},\ \bibinfo {pages}
  {022001} (\bibinfo {year} {2017})}\BibitemShut {NoStop}%
\bibitem [{\citenamefont {Collopy}\ \emph {et~al.}(2015)\citenamefont
  {Collopy}, \citenamefont {Hummon}, \citenamefont {Yeo}, \citenamefont {Yan},\
  and\ \citenamefont {Ye}}]{Ye}%
  \BibitemOpen
  \bibfield  {author} {\bibinfo {author} {\bibfnamefont {A.~L.}\ \bibnamefont
  {Collopy}}, \bibinfo {author} {\bibfnamefont {M.~T.}\ \bibnamefont {Hummon}},
  \bibinfo {author} {\bibfnamefont {M.}~\bibnamefont {Yeo}}, \bibinfo {author}
  {\bibfnamefont {B.}~\bibnamefont {Yan}}, \ and\ \bibinfo {author}
  {\bibfnamefont {J.}~\bibnamefont {Ye}},\ }\href {\doibase
  10.1088/1367-2630/17/5/055008} {\bibfield  {journal} {\bibinfo  {journal}
  {New Journal of Physics}\ }\textbf {\bibinfo {volume} {17}},\ \bibinfo
  {pages} {055008} (\bibinfo {year} {2015})}\BibitemShut {NoStop}%
\bibitem [{\citenamefont {Hemmerling}\ \emph {et~al.}(2016)\citenamefont
  {Hemmerling}, \citenamefont {Chae}, \citenamefont {Ravi}, \citenamefont
  {Anderegg}, \citenamefont {Drayna}, \citenamefont {Hutzler}, \citenamefont
  {Collopy}, \citenamefont {Ye}, \citenamefont {Ketterle},\ and\ \citenamefont
  {Doyle}}]{Hemmerling}%
  \BibitemOpen
  \bibfield  {author} {\bibinfo {author} {\bibfnamefont {B.}~\bibnamefont
  {Hemmerling}}, \bibinfo {author} {\bibfnamefont {E.}~\bibnamefont {Chae}},
  \bibinfo {author} {\bibfnamefont {A.}~\bibnamefont {Ravi}}, \bibinfo {author}
  {\bibfnamefont {L.}~\bibnamefont {Anderegg}}, \bibinfo {author}
  {\bibfnamefont {G.~K.}\ \bibnamefont {Drayna}}, \bibinfo {author}
  {\bibfnamefont {N.~R.}\ \bibnamefont {Hutzler}}, \bibinfo {author}
  {\bibfnamefont {A.~L.}\ \bibnamefont {Collopy}}, \bibinfo {author}
  {\bibfnamefont {J.}~\bibnamefont {Ye}}, \bibinfo {author} {\bibfnamefont
  {W.}~\bibnamefont {Ketterle}}, \ and\ \bibinfo {author} {\bibfnamefont
  {J.~M.}\ \bibnamefont {Doyle}},\ }\href {\doibase
  10.1088/0953-4075/49/17/174001} {\bibfield  {journal} {\bibinfo  {journal}
  {Journal of Physics B: Atomic, Molecular and Optical Physics}\ }\textbf
  {\bibinfo {volume} {49}},\ \bibinfo {pages} {174001} (\bibinfo {year}
  {2016})}\BibitemShut {NoStop}%
\bibitem [{\citenamefont {{Lunden}}\ \emph {et~al.}(2019)\citenamefont
  {{Lunden}}, \citenamefont {{Du}}, \citenamefont {{Cantara}}, \citenamefont
  {{Barral}}, \citenamefont {{Jamison}},\ and\ \citenamefont
  {{Ketterle}}}]{Lunden}%
  \BibitemOpen
  \bibfield  {author} {\bibinfo {author} {\bibfnamefont {W.}~\bibnamefont
  {{Lunden}}}, \bibinfo {author} {\bibfnamefont {L.}~\bibnamefont {{Du}}},
  \bibinfo {author} {\bibfnamefont {M.}~\bibnamefont {{Cantara}}}, \bibinfo
  {author} {\bibfnamefont {P.}~\bibnamefont {{Barral}}}, \bibinfo {author}
  {\bibfnamefont {A.~O.}\ \bibnamefont {{Jamison}}}, \ and\ \bibinfo {author}
  {\bibfnamefont {W.}~\bibnamefont {{Ketterle}}},\ }\href@noop {} {\bibfield
  {journal} {\bibinfo  {journal} {arXiv e-prints}\ ,\ \bibinfo {eid}
  {arXiv:1908.10433}} (\bibinfo {year} {2019})},\ \Eprint
  {http://arxiv.org/abs/1908.10433} {arXiv:1908.10433 [cond-mat.quant-gas]}
  \BibitemShut {NoStop}%
\bibitem [{\citenamefont {Stellmer}\ \emph {et~al.}(2013)\citenamefont
  {Stellmer}, \citenamefont {Pasquiou}, \citenamefont {Grimm},\ and\
  \citenamefont {Schreck}}]{Stellmer}%
  \BibitemOpen
  \bibfield  {author} {\bibinfo {author} {\bibfnamefont {S.}~\bibnamefont
  {Stellmer}}, \bibinfo {author} {\bibfnamefont {B.}~\bibnamefont {Pasquiou}},
  \bibinfo {author} {\bibfnamefont {R.}~\bibnamefont {Grimm}}, \ and\ \bibinfo
  {author} {\bibfnamefont {F.}~\bibnamefont {Schreck}},\ }\href {\doibase
  10.1103/PhysRevLett.110.263003} {\bibfield  {journal} {\bibinfo  {journal}
  {Phys. Rev. Lett.}\ }\textbf {\bibinfo {volume} {110}},\ \bibinfo {pages}
  {263003} (\bibinfo {year} {2013})}\BibitemShut {NoStop}%
\bibitem [{\citenamefont {Urvoy}\ \emph {et~al.}(2019)\citenamefont {Urvoy},
  \citenamefont {Vendeiro}, \citenamefont {Ramette}, \citenamefont
  {Adiyatullin},\ and\ \citenamefont {Vuleti\ifmmode~\acute{c}\else
  \'{c}\fi{}}}]{Urvoy}%
  \BibitemOpen
  \bibfield  {author} {\bibinfo {author} {\bibfnamefont {A.}~\bibnamefont
  {Urvoy}}, \bibinfo {author} {\bibfnamefont {Z.}~\bibnamefont {Vendeiro}},
  \bibinfo {author} {\bibfnamefont {J.}~\bibnamefont {Ramette}}, \bibinfo
  {author} {\bibfnamefont {A.}~\bibnamefont {Adiyatullin}}, \ and\ \bibinfo
  {author} {\bibfnamefont {V.}~\bibnamefont {Vuleti\ifmmode~\acute{c}\else
  \'{c}\fi{}}},\ }\href {\doibase 10.1103/PhysRevLett.122.203202} {\bibfield
  {journal} {\bibinfo  {journal} {Phys. Rev. Lett.}\ }\textbf {\bibinfo
  {volume} {122}},\ \bibinfo {pages} {203202} (\bibinfo {year}
  {2019})}\BibitemShut {NoStop}%
\bibitem [{\citenamefont {Stuhl}\ \emph {et~al.}(2012)\citenamefont {Stuhl},
  \citenamefont {Hummon}, \citenamefont {Yeo}, \citenamefont {Quéméner},
  \citenamefont {Bohn},\ and\ \citenamefont {Ye}}]{Stuhl}%
  \BibitemOpen
  \bibfield  {author} {\bibinfo {author} {\bibfnamefont {B.~K.}\ \bibnamefont
  {Stuhl}}, \bibinfo {author} {\bibfnamefont {M.~T.}\ \bibnamefont {Hummon}},
  \bibinfo {author} {\bibfnamefont {M.}~\bibnamefont {Yeo}}, \bibinfo {author}
  {\bibfnamefont {G.}~\bibnamefont {Quéméner}}, \bibinfo {author}
  {\bibfnamefont {J.~L.}\ \bibnamefont {Bohn}}, \ and\ \bibinfo {author}
  {\bibfnamefont {J.}~\bibnamefont {Ye}},\ }\href {\doibase
  10.1038/nature11718} {\bibfield  {journal} {\bibinfo  {journal} {Nature}\
  }\textbf {\bibinfo {volume} {492}},\ \bibinfo {pages} {396} (\bibinfo {year}
  {2012})}\BibitemShut {NoStop}%
\bibitem [{\citenamefont {Elliott}\ \emph {et~al.}(2018)\citenamefont
  {Elliott}, \citenamefont {Krutzik}, \citenamefont {Williams}, \citenamefont
  {Thompson},\ and\ \citenamefont {Aveline}}]{Elliott}%
  \BibitemOpen
  \bibfield  {author} {\bibinfo {author} {\bibfnamefont {E.~R.}\ \bibnamefont
  {Elliott}}, \bibinfo {author} {\bibfnamefont {M.~C.}\ \bibnamefont
  {Krutzik}}, \bibinfo {author} {\bibfnamefont {J.~R.}\ \bibnamefont
  {Williams}}, \bibinfo {author} {\bibfnamefont {R.~J.}\ \bibnamefont
  {Thompson}}, \ and\ \bibinfo {author} {\bibfnamefont {D.~C.}\ \bibnamefont
  {Aveline}},\ }\href {\doibase 10.1038/s41526-018-0049-9} {\bibfield
  {journal} {\bibinfo  {journal} {npj Microgravity}\ }\textbf {\bibinfo
  {volume} {4}},\ \bibinfo {pages} {16} (\bibinfo {year} {2018})}\BibitemShut
  {NoStop}%
\bibitem [{\citenamefont {Norcia}\ \emph {et~al.}(2018)\citenamefont {Norcia},
  \citenamefont {Cline}, \citenamefont {Bartolotta}, \citenamefont {Holland},\
  and\ \citenamefont {Thompson}}]{SWAPExperiment}%
  \BibitemOpen
  \bibfield  {author} {\bibinfo {author} {\bibfnamefont {M.~A.}\ \bibnamefont
  {Norcia}}, \bibinfo {author} {\bibfnamefont {J.~R.~K.}\ \bibnamefont
  {Cline}}, \bibinfo {author} {\bibfnamefont {J.~P.}\ \bibnamefont
  {Bartolotta}}, \bibinfo {author} {\bibfnamefont {M.~J.}\ \bibnamefont
  {Holland}}, \ and\ \bibinfo {author} {\bibfnamefont {J.~K.}\ \bibnamefont
  {Thompson}},\ }\href {http://stacks.iop.org/1367-2630/20/i=2/a=023021}
  {\bibfield  {journal} {\bibinfo  {journal} {New Journal of Physics}\ }\textbf
  {\bibinfo {volume} {20}},\ \bibinfo {pages} {023021} (\bibinfo {year}
  {2018})}\BibitemShut {NoStop}%
\bibitem [{\citenamefont {Bartolotta}\ \emph {et~al.}(2018)\citenamefont
  {Bartolotta}, \citenamefont {Norcia}, \citenamefont {Cline}, \citenamefont
  {Thompson},\ and\ \citenamefont {Holland}}]{SWAPTheory}%
  \BibitemOpen
  \bibfield  {author} {\bibinfo {author} {\bibfnamefont {J.~P.}\ \bibnamefont
  {Bartolotta}}, \bibinfo {author} {\bibfnamefont {M.~A.}\ \bibnamefont
  {Norcia}}, \bibinfo {author} {\bibfnamefont {J.~R.~K.}\ \bibnamefont
  {Cline}}, \bibinfo {author} {\bibfnamefont {J.~K.}\ \bibnamefont {Thompson}},
  \ and\ \bibinfo {author} {\bibfnamefont {M.~J.}\ \bibnamefont {Holland}},\
  }\href {\doibase 10.1103/physreva.98.023404} {\bibfield  {journal} {\bibinfo
  {journal} {Physical Review A}\ }\textbf {\bibinfo {volume} {98}} (\bibinfo
  {year} {2018}),\ 10.1103/physreva.98.023404}\BibitemShut {NoStop}%
\bibitem [{\citenamefont {Greve}\ \emph {et~al.}(2019)\citenamefont {Greve},
  \citenamefont {Wu},\ and\ \citenamefont {Thompson}}]{SWAPRaman}%
  \BibitemOpen
  \bibfield  {author} {\bibinfo {author} {\bibfnamefont {G.~P.}\ \bibnamefont
  {Greve}}, \bibinfo {author} {\bibfnamefont {B.}~\bibnamefont {Wu}}, \ and\
  \bibinfo {author} {\bibfnamefont {J.~K.}\ \bibnamefont {Thompson}},\ }\href
  {\doibase 10.1088/1367-2630/ab2f3c} {\bibfield  {journal} {\bibinfo
  {journal} {New Journal of Physics}\ }\textbf {\bibinfo {volume} {21}},\
  \bibinfo {pages} {073045} (\bibinfo {year} {2019})}\BibitemShut {NoStop}%
\bibitem [{\citenamefont {Bartolotta}\ and\ \citenamefont
  {Holland}(2020)}]{SWAP_MOT_theory}%
  \BibitemOpen
  \bibfield  {author} {\bibinfo {author} {\bibfnamefont {J.~P.}\ \bibnamefont
  {Bartolotta}}\ and\ \bibinfo {author} {\bibfnamefont {M.~J.}\ \bibnamefont
  {Holland}},\ }\href {\doibase 10.1103/PhysRevA.101.053434} {\bibfield
  {journal} {\bibinfo  {journal} {Phys. Rev. A}\ }\textbf {\bibinfo {volume}
  {101}},\ \bibinfo {pages} {053434} (\bibinfo {year} {2020})}\BibitemShut
  {NoStop}%
\bibitem [{\citenamefont {Snigirev}\ \emph {et~al.}(2019)\citenamefont
  {Snigirev}, \citenamefont {Park}, \citenamefont {Heinz}, \citenamefont
  {Bloch},\ and\ \citenamefont {Blatt}}]{Snigirev}%
  \BibitemOpen
  \bibfield  {author} {\bibinfo {author} {\bibfnamefont {S.}~\bibnamefont
  {Snigirev}}, \bibinfo {author} {\bibfnamefont {A.~J.}\ \bibnamefont {Park}},
  \bibinfo {author} {\bibfnamefont {A.}~\bibnamefont {Heinz}}, \bibinfo
  {author} {\bibfnamefont {I.}~\bibnamefont {Bloch}}, \ and\ \bibinfo {author}
  {\bibfnamefont {S.}~\bibnamefont {Blatt}},\ }\href {\doibase
  10.1103/PhysRevA.99.063421} {\bibfield  {journal} {\bibinfo  {journal} {Phys.
  Rev. A}\ }\textbf {\bibinfo {volume} {99}},\ \bibinfo {pages} {063421}
  (\bibinfo {year} {2019})}\BibitemShut {NoStop}%
\bibitem [{\citenamefont {{Muniz}}\ \emph {et~al.}(2018)\citenamefont
  {{Muniz}}, \citenamefont {{Norcia}}, \citenamefont {{Cline}},\ and\
  \citenamefont {{Thompson}}}]{muniz}%
  \BibitemOpen
  \bibfield  {author} {\bibinfo {author} {\bibfnamefont {J.~A.}\ \bibnamefont
  {{Muniz}}}, \bibinfo {author} {\bibfnamefont {M.~A.}\ \bibnamefont
  {{Norcia}}}, \bibinfo {author} {\bibfnamefont {J.~R.~K.}\ \bibnamefont
  {{Cline}}}, \ and\ \bibinfo {author} {\bibfnamefont {J.~K.}\ \bibnamefont
  {{Thompson}}},\ }\href@noop {} {\bibfield  {journal} {\bibinfo  {journal}
  {arXiv e-prints}\ ,\ \bibinfo {eid} {arXiv:1806.00838}} (\bibinfo {year}
  {2018})},\ \Eprint {http://arxiv.org/abs/1806.00838} {arXiv:1806.00838
  [physics.atom-ph]} \BibitemShut {NoStop}%
\bibitem [{\citenamefont {Allen}\ and\ \citenamefont {Eberly}(1987)}]{AE}%
  \BibitemOpen
  \bibfield  {author} {\bibinfo {author} {\bibfnamefont {L.}~\bibnamefont
  {Allen}}\ and\ \bibinfo {author} {\bibfnamefont {J.~H.}\ \bibnamefont
  {Eberly}},\ }\href@noop {} {\emph {\bibinfo {title} {Optical Resonance and
  Two-level Atoms}}}\ (\bibinfo  {publisher} {Dover},\ \bibinfo {year}
  {1987})\BibitemShut {NoStop}%
\bibitem [{\citenamefont {Voitsekhovich}\ \emph {et~al.}(1991)\citenamefont
  {Voitsekhovich}, \citenamefont {Danileiko}, \citenamefont {Negriiko},
  \citenamefont {Romanenko},\ and\ \citenamefont {Yatsenko}}]{Voitsekhovich}%
  \BibitemOpen
  \bibfield  {author} {\bibinfo {author} {\bibfnamefont {V.}~\bibnamefont
  {Voitsekhovich}}, \bibinfo {author} {\bibfnamefont {M.}~\bibnamefont
  {Danileiko}}, \bibinfo {author} {\bibfnamefont {A.}~\bibnamefont {Negriiko}},
  \bibinfo {author} {\bibfnamefont {V.}~\bibnamefont {Romanenko}}, \ and\
  \bibinfo {author} {\bibfnamefont {L.}~\bibnamefont {Yatsenko}},\ }\href
  {http://www.jetp.ac.ru/cgi-bin/dn/e_072_02_0219.pdf} {\bibfield  {journal}
  {\bibinfo  {journal} {JETP}\ }\textbf {\bibinfo {volume} {72}},\ \bibinfo
  {pages} {219} (\bibinfo {year} {1991})}\BibitemShut {NoStop}%
\bibitem [{\citenamefont {S\"oding}\ \emph {et~al.}(1997)\citenamefont
  {S\"oding}, \citenamefont {Grimm}, \citenamefont {Ovchinnikov}, \citenamefont
  {Bouyer},\ and\ \citenamefont {Salomon}}]{SalomonBCF}%
  \BibitemOpen
  \bibfield  {author} {\bibinfo {author} {\bibfnamefont {J.}~\bibnamefont
  {S\"oding}}, \bibinfo {author} {\bibfnamefont {R.}~\bibnamefont {Grimm}},
  \bibinfo {author} {\bibfnamefont {Y.~B.}\ \bibnamefont {Ovchinnikov}},
  \bibinfo {author} {\bibfnamefont {P.}~\bibnamefont {Bouyer}}, \ and\ \bibinfo
  {author} {\bibfnamefont {C.}~\bibnamefont {Salomon}},\ }\href {\doibase
  10.1103/PhysRevLett.78.1420} {\bibfield  {journal} {\bibinfo  {journal}
  {Phys. Rev. Lett.}\ }\textbf {\bibinfo {volume} {78}},\ \bibinfo {pages}
  {1420} (\bibinfo {year} {1997})}\BibitemShut {NoStop}%
\bibitem [{\citenamefont {Cashen}\ and\ \citenamefont
  {Metcalf}(2001)}]{MetcalfBCF}%
  \BibitemOpen
  \bibfield  {author} {\bibinfo {author} {\bibfnamefont {M.~T.}\ \bibnamefont
  {Cashen}}\ and\ \bibinfo {author} {\bibfnamefont {H.}~\bibnamefont
  {Metcalf}},\ }\href {\doibase 10.1103/PhysRevA.63.025406} {\bibfield
  {journal} {\bibinfo  {journal} {Phys. Rev. A}\ }\textbf {\bibinfo {volume}
  {63}},\ \bibinfo {pages} {025406} (\bibinfo {year} {2001})}\BibitemShut
  {NoStop}%
\bibitem [{\citenamefont {Miao}\ \emph {et~al.}(2007)\citenamefont {Miao},
  \citenamefont {Wertz}, \citenamefont {Cohen},\ and\ \citenamefont
  {Metcalf}}]{ARP1}%
  \BibitemOpen
  \bibfield  {author} {\bibinfo {author} {\bibfnamefont {X.}~\bibnamefont
  {Miao}}, \bibinfo {author} {\bibfnamefont {E.}~\bibnamefont {Wertz}},
  \bibinfo {author} {\bibfnamefont {M.~G.}\ \bibnamefont {Cohen}}, \ and\
  \bibinfo {author} {\bibfnamefont {H.}~\bibnamefont {Metcalf}},\ }\href
  {\doibase 10.1103/PhysRevA.75.011402} {\bibfield  {journal} {\bibinfo
  {journal} {Phys. Rev. A}\ }\textbf {\bibinfo {volume} {75}},\ \bibinfo
  {pages} {011402} (\bibinfo {year} {2007})}\BibitemShut {NoStop}%
\bibitem [{\citenamefont {Stack}\ \emph {et~al.}(2011)\citenamefont {Stack},
  \citenamefont {Elgin}, \citenamefont {Anisimov},\ and\ \citenamefont
  {Metcalf}}]{ARP2}%
  \BibitemOpen
  \bibfield  {author} {\bibinfo {author} {\bibfnamefont {D.}~\bibnamefont
  {Stack}}, \bibinfo {author} {\bibfnamefont {J.}~\bibnamefont {Elgin}},
  \bibinfo {author} {\bibfnamefont {P.~M.}\ \bibnamefont {Anisimov}}, \ and\
  \bibinfo {author} {\bibfnamefont {H.}~\bibnamefont {Metcalf}},\ }\href
  {\doibase 10.1103/PhysRevA.84.013420} {\bibfield  {journal} {\bibinfo
  {journal} {Phys. Rev. A}\ }\textbf {\bibinfo {volume} {84}},\ \bibinfo
  {pages} {013420} (\bibinfo {year} {2011})}\BibitemShut {NoStop}%
\bibitem [{\citenamefont {Metcalf}(2017)}]{MetcalfColloquium}%
  \BibitemOpen
  \bibfield  {author} {\bibinfo {author} {\bibfnamefont {H.}~\bibnamefont
  {Metcalf}},\ }\href {\doibase 10.1103/RevModPhys.89.041001} {\bibfield
  {journal} {\bibinfo  {journal} {Rev. Mod. Phys.}\ }\textbf {\bibinfo {volume}
  {89}},\ \bibinfo {pages} {041001} (\bibinfo {year} {2017})}\BibitemShut
  {NoStop}%
\bibitem [{\citenamefont {Gu\'ery-Odelin}\ \emph {et~al.}(2019)\citenamefont
  {Gu\'ery-Odelin}, \citenamefont {Ruschhaupt}, \citenamefont {Kiely},
  \citenamefont {Torrontegui}, \citenamefont {Mart\'{\i}nez-Garaot},\ and\
  \citenamefont {Muga}}]{Review}%
  \BibitemOpen
  \bibfield  {author} {\bibinfo {author} {\bibfnamefont {D.}~\bibnamefont
  {Gu\'ery-Odelin}}, \bibinfo {author} {\bibfnamefont {A.}~\bibnamefont
  {Ruschhaupt}}, \bibinfo {author} {\bibfnamefont {A.}~\bibnamefont {Kiely}},
  \bibinfo {author} {\bibfnamefont {E.}~\bibnamefont {Torrontegui}}, \bibinfo
  {author} {\bibfnamefont {S.}~\bibnamefont {Mart\'{\i}nez-Garaot}}, \ and\
  \bibinfo {author} {\bibfnamefont {J.~G.}\ \bibnamefont {Muga}},\ }\href
  {\doibase 10.1103/RevModPhys.91.045001} {\bibfield  {journal} {\bibinfo
  {journal} {Rev. Mod. Phys.}\ }\textbf {\bibinfo {volume} {91}},\ \bibinfo
  {pages} {045001} (\bibinfo {year} {2019})}\BibitemShut {NoStop}%
\bibitem [{\citenamefont {Lewis}\ and\ \citenamefont {Riesenfeld}(1969)}]{LRI}%
  \BibitemOpen
  \bibfield  {author} {\bibinfo {author} {\bibfnamefont {H.~R.}\ \bibnamefont
  {Lewis}}\ and\ \bibinfo {author} {\bibfnamefont {W.~B.}\ \bibnamefont
  {Riesenfeld}},\ }\href {\doibase 10.1063/1.1664991} {\bibfield  {journal}
  {\bibinfo  {journal} {Journal of Mathematical Physics}\ }\textbf {\bibinfo
  {volume} {10}},\ \bibinfo {pages} {1458} (\bibinfo {year} {1969})},\ \Eprint
  {http://arxiv.org/abs/https://doi.org/10.1063/1.1664991}
  {https://doi.org/10.1063/1.1664991} \BibitemShut {NoStop}%
\bibitem [{\citenamefont {Chen}\ \emph {et~al.}(2011)\citenamefont {Chen},
  \citenamefont {Torrontegui},\ and\ \citenamefont {Muga}}]{LRIandTQD}%
  \BibitemOpen
  \bibfield  {author} {\bibinfo {author} {\bibfnamefont {X.}~\bibnamefont
  {Chen}}, \bibinfo {author} {\bibfnamefont {E.}~\bibnamefont {Torrontegui}}, \
  and\ \bibinfo {author} {\bibfnamefont {J.~G.}\ \bibnamefont {Muga}},\ }\href
  {\doibase 10.1103/PhysRevA.83.062116} {\bibfield  {journal} {\bibinfo
  {journal} {Phys. Rev. A}\ }\textbf {\bibinfo {volume} {83}},\ \bibinfo
  {pages} {062116} (\bibinfo {year} {2011})}\BibitemShut {NoStop}%
\bibitem [{\citenamefont {Chen}\ \emph {et~al.}(2010)\citenamefont {Chen},
  \citenamefont {Ruschhaupt}, \citenamefont {Schmidt}, \citenamefont {del
  Campo}, \citenamefont {Gu\'ery-Odelin},\ and\ \citenamefont
  {Muga}}]{LRI_PRL}%
  \BibitemOpen
  \bibfield  {author} {\bibinfo {author} {\bibfnamefont {X.}~\bibnamefont
  {Chen}}, \bibinfo {author} {\bibfnamefont {A.}~\bibnamefont {Ruschhaupt}},
  \bibinfo {author} {\bibfnamefont {S.}~\bibnamefont {Schmidt}}, \bibinfo
  {author} {\bibfnamefont {A.}~\bibnamefont {del Campo}}, \bibinfo {author}
  {\bibfnamefont {D.}~\bibnamefont {Gu\'ery-Odelin}}, \ and\ \bibinfo {author}
  {\bibfnamefont {J.~G.}\ \bibnamefont {Muga}},\ }\href {\doibase
  10.1103/PhysRevLett.104.063002} {\bibfield  {journal} {\bibinfo  {journal}
  {Phys. Rev. Lett.}\ }\textbf {\bibinfo {volume} {104}},\ \bibinfo {pages}
  {063002} (\bibinfo {year} {2010})}\BibitemShut {NoStop}%
\bibitem [{\citenamefont {Lai}\ \emph {et~al.}(1996)\citenamefont {Lai},
  \citenamefont {Liang}, \citenamefont {M\"uller-Kirsten},\ and\ \citenamefont
  {Zhou}}]{Lai}%
  \BibitemOpen
  \bibfield  {author} {\bibinfo {author} {\bibfnamefont {Y.-Z.}\ \bibnamefont
  {Lai}}, \bibinfo {author} {\bibfnamefont {J.-Q.}\ \bibnamefont {Liang}},
  \bibinfo {author} {\bibfnamefont {H.~J.~W.}\ \bibnamefont
  {M\"uller-Kirsten}}, \ and\ \bibinfo {author} {\bibfnamefont {J.-G.}\
  \bibnamefont {Zhou}},\ }\href {\doibase 10.1103/PhysRevA.53.3691} {\bibfield
  {journal} {\bibinfo  {journal} {Phys. Rev. A}\ }\textbf {\bibinfo {volume}
  {53}},\ \bibinfo {pages} {3691} (\bibinfo {year} {1996})}\BibitemShut
  {NoStop}%
\bibitem [{\citenamefont {Chen}\ \emph {et~al.}(2014)\citenamefont {Chen},
  \citenamefont {Xia}, \citenamefont {Chen},\ and\ \citenamefont
  {Song}}]{Chen_2014}%
  \BibitemOpen
  \bibfield  {author} {\bibinfo {author} {\bibfnamefont {Y.-H.}\ \bibnamefont
  {Chen}}, \bibinfo {author} {\bibfnamefont {Y.}~\bibnamefont {Xia}}, \bibinfo
  {author} {\bibfnamefont {Q.-Q.}\ \bibnamefont {Chen}}, \ and\ \bibinfo
  {author} {\bibfnamefont {J.}~\bibnamefont {Song}},\ }\href {\doibase
  10.1088/1612-2011/11/11/115201} {\bibfield  {journal} {\bibinfo  {journal}
  {Laser Physics Letters}\ }\textbf {\bibinfo {volume} {11}},\ \bibinfo {pages}
  {115201} (\bibinfo {year} {2014})}\BibitemShut {NoStop}%
\bibitem [{\citenamefont {G\"ung\"ord\"u}\ \emph {et~al.}(2012)\citenamefont
  {G\"ung\"ord\"u}, \citenamefont {Wan}, \citenamefont {Fasihi},\ and\
  \citenamefont {Nakahara}}]{LRIFourLevel}%
  \BibitemOpen
  \bibfield  {author} {\bibinfo {author} {\bibfnamefont {U.}~\bibnamefont
  {G\"ung\"ord\"u}}, \bibinfo {author} {\bibfnamefont {Y.}~\bibnamefont {Wan}},
  \bibinfo {author} {\bibfnamefont {M.~A.}\ \bibnamefont {Fasihi}}, \ and\
  \bibinfo {author} {\bibfnamefont {M.}~\bibnamefont {Nakahara}},\ }\href
  {\doibase 10.1103/PhysRevA.86.062312} {\bibfield  {journal} {\bibinfo
  {journal} {Phys. Rev. A}\ }\textbf {\bibinfo {volume} {86}},\ \bibinfo
  {pages} {062312} (\bibinfo {year} {2012})}\BibitemShut {NoStop}%
\bibitem [{\citenamefont {Zener}(1932)}]{zener}%
  \BibitemOpen
  \bibfield  {author} {\bibinfo {author} {\bibfnamefont {C.}~\bibnamefont
  {Zener}},\ }\href {\doibase 10.1098/rspa.1932.0165} {\bibfield  {journal}
  {\bibinfo  {journal} {Proceedings of the Royal Society of London. Series A,
  Containing Papers of a Mathematical and Physical Character}\ }\textbf
  {\bibinfo {volume} {137}},\ \bibinfo {pages} {696–702} (\bibinfo {year}
  {1932})}\BibitemShut {NoStop}%
\bibitem [{\citenamefont {Camparo}\ and\ \citenamefont
  {Frueholz}(1984)}]{Camparo_1984}%
  \BibitemOpen
  \bibfield  {author} {\bibinfo {author} {\bibfnamefont {J.~C.}\ \bibnamefont
  {Camparo}}\ and\ \bibinfo {author} {\bibfnamefont {R.~P.}\ \bibnamefont
  {Frueholz}},\ }\href {\doibase 10.1088/0022-3700/17/20/015} {\bibfield
  {journal} {\bibinfo  {journal} {Journal of Physics B: Atomic and Molecular
  Physics}\ }\textbf {\bibinfo {volume} {17}},\ \bibinfo {pages} {4169}
  (\bibinfo {year} {1984})}\BibitemShut {NoStop}%
\bibitem [{\citenamefont {Feynman}\ \emph {et~al.}(1957)\citenamefont
  {Feynman}, \citenamefont {Vernon},\ and\ \citenamefont
  {Hellwarth}}]{feynman_vernon_hellwarth_1957}%
  \BibitemOpen
  \bibfield  {author} {\bibinfo {author} {\bibfnamefont {R.~P.}\ \bibnamefont
  {Feynman}}, \bibinfo {author} {\bibfnamefont {F.~L.}\ \bibnamefont {Vernon}},
  \ and\ \bibinfo {author} {\bibfnamefont {R.~W.}\ \bibnamefont {Hellwarth}},\
  }\href {\doibase 10.1063/1.1722572} {\bibfield  {journal} {\bibinfo
  {journal} {Journal of Applied Physics}\ }\textbf {\bibinfo {volume} {28}},\
  \bibinfo {pages} {49–52} (\bibinfo {year} {1957})}\BibitemShut {NoStop}%
\bibitem [{\citenamefont {Vitanov}(1999)}]{vitanov}%
  \BibitemOpen
  \bibfield  {author} {\bibinfo {author} {\bibfnamefont {N.~V.}\ \bibnamefont
  {Vitanov}},\ }\href {\doibase 10.1103/PhysRevA.59.988} {\bibfield  {journal}
  {\bibinfo  {journal} {Phys. Rev. A}\ }\textbf {\bibinfo {volume} {59}},\
  \bibinfo {pages} {988} (\bibinfo {year} {1999})}\BibitemShut {NoStop}%
\bibitem [{\citenamefont {Sawicki}\ and\ \citenamefont
  {Eberly}(1999)}]{eberly}%
  \BibitemOpen
  \bibfield  {author} {\bibinfo {author} {\bibfnamefont {D.}~\bibnamefont
  {Sawicki}}\ and\ \bibinfo {author} {\bibfnamefont {J.}~\bibnamefont
  {Eberly}},\ }\href {\doibase 10.1364/OE.4.000217} {\bibfield  {journal}
  {\bibinfo  {journal} {Opt. Express}\ }\textbf {\bibinfo {volume} {4}},\
  \bibinfo {pages} {217} (\bibinfo {year} {1999})}\BibitemShut {NoStop}%
\bibitem [{\citenamefont {M{\o}lmer}\ \emph {et~al.}(1993)\citenamefont
  {M{\o}lmer}, \citenamefont {Castin},\ and\ \citenamefont
  {Dalibard}}]{Molmer}%
  \BibitemOpen
  \bibfield  {author} {\bibinfo {author} {\bibfnamefont {K.}~\bibnamefont
  {M{\o}lmer}}, \bibinfo {author} {\bibfnamefont {Y.}~\bibnamefont {Castin}}, \
  and\ \bibinfo {author} {\bibfnamefont {J.}~\bibnamefont {Dalibard}},\ }\href
  {\doibase 10.1364/JOSAB.10.000524} {\bibfield  {journal} {\bibinfo  {journal}
  {J. Opt. Soc. Am. B}\ }\textbf {\bibinfo {volume} {10}},\ \bibinfo {pages}
  {524} (\bibinfo {year} {1993})}\BibitemShut {NoStop}%
\bibitem [{\citenamefont {Minogin}\ and\ \citenamefont
  {Serimaa}(1979)}]{dopplerons}%
  \BibitemOpen
  \bibfield  {author} {\bibinfo {author} {\bibfnamefont {V.}~\bibnamefont
  {Minogin}}\ and\ \bibinfo {author} {\bibfnamefont {O.}~\bibnamefont
  {Serimaa}},\ }\href {\doibase 10.1016/0030-4018(79)90374-2} {\bibfield
  {journal} {\bibinfo  {journal} {Optics Communications}\ }\textbf {\bibinfo
  {volume} {30}},\ \bibinfo {pages} {373–379} (\bibinfo {year}
  {1979})}\BibitemShut {NoStop}%
\bibitem [{\citenamefont {Stefanatos}\ \emph {et~al.}(2010)\citenamefont
  {Stefanatos}, \citenamefont {Ruths},\ and\ \citenamefont {Li}}]{Stefanatos}%
  \BibitemOpen
  \bibfield  {author} {\bibinfo {author} {\bibfnamefont {D.}~\bibnamefont
  {Stefanatos}}, \bibinfo {author} {\bibfnamefont {J.}~\bibnamefont {Ruths}}, \
  and\ \bibinfo {author} {\bibfnamefont {J.-S.}\ \bibnamefont {Li}},\ }\href
  {\doibase 10.1103/PhysRevA.82.063422} {\bibfield  {journal} {\bibinfo
  {journal} {Phys. Rev. A}\ }\textbf {\bibinfo {volume} {82}},\ \bibinfo
  {pages} {063422} (\bibinfo {year} {2010})}\BibitemShut {NoStop}%
\bibitem [{\citenamefont {Ruschhaupt}\ and\ \citenamefont
  {Muga}(2014)}]{Control}%
  \BibitemOpen
  \bibfield  {author} {\bibinfo {author} {\bibfnamefont {A.}~\bibnamefont
  {Ruschhaupt}}\ and\ \bibinfo {author} {\bibfnamefont {J.}~\bibnamefont
  {Muga}},\ }\href {\doibase 10.1080/09500340.2013.846431} {\bibfield
  {journal} {\bibinfo  {journal} {Journal of Modern Optics}\ }\textbf {\bibinfo
  {volume} {61}},\ \bibinfo {pages} {828} (\bibinfo {year} {2014})},\ \Eprint
  {http://arxiv.org/abs/https://doi.org/10.1080/09500340.2013.846431}
  {https://doi.org/10.1080/09500340.2013.846431} \BibitemShut {NoStop}%
\bibitem [{\citenamefont {Mortensen}\ \emph {et~al.}(2018)\citenamefont
  {Mortensen}, \citenamefont {S{\o}rensen}, \citenamefont {M{\o}lmer},\ and\
  \citenamefont {Sherson}}]{Mortensen}%
  \BibitemOpen
  \bibfield  {author} {\bibinfo {author} {\bibfnamefont {H.~L.}\ \bibnamefont
  {Mortensen}}, \bibinfo {author} {\bibfnamefont {J.~J. W.~H.}\ \bibnamefont
  {S{\o}rensen}}, \bibinfo {author} {\bibfnamefont {K.}~\bibnamefont
  {M{\o}lmer}}, \ and\ \bibinfo {author} {\bibfnamefont {J.~F.}\ \bibnamefont
  {Sherson}},\ }\href {\doibase 10.1088/1367-2630/aaac8a} {\bibfield  {journal}
  {\bibinfo  {journal} {New Journal of Physics}\ }\textbf {\bibinfo {volume}
  {20}},\ \bibinfo {pages} {025009} (\bibinfo {year} {2018})}\BibitemShut
  {NoStop}%
\bibitem [{\citenamefont {Campbell}\ and\ \citenamefont
  {Deffner}(2017)}]{Campbell}%
  \BibitemOpen
  \bibfield  {author} {\bibinfo {author} {\bibfnamefont {S.}~\bibnamefont
  {Campbell}}\ and\ \bibinfo {author} {\bibfnamefont {S.}~\bibnamefont
  {Deffner}},\ }\href {\doibase 10.1103/PhysRevLett.118.100601} {\bibfield
  {journal} {\bibinfo  {journal} {Phys. Rev. Lett.}\ }\textbf {\bibinfo
  {volume} {118}},\ \bibinfo {pages} {100601} (\bibinfo {year}
  {2017})}\BibitemShut {NoStop}%
\bibitem [{\citenamefont {Abah}\ \emph {et~al.}(2019)\citenamefont {Abah},
  \citenamefont {Puebla}, \citenamefont {Kiely}, \citenamefont {Chiara},
  \citenamefont {Paternostro},\ and\ \citenamefont {Campbell}}]{Abah}%
  \BibitemOpen
  \bibfield  {author} {\bibinfo {author} {\bibfnamefont {O.}~\bibnamefont
  {Abah}}, \bibinfo {author} {\bibfnamefont {R.}~\bibnamefont {Puebla}},
  \bibinfo {author} {\bibfnamefont {A.}~\bibnamefont {Kiely}}, \bibinfo
  {author} {\bibfnamefont {G.~D.}\ \bibnamefont {Chiara}}, \bibinfo {author}
  {\bibfnamefont {M.}~\bibnamefont {Paternostro}}, \ and\ \bibinfo {author}
  {\bibfnamefont {S.}~\bibnamefont {Campbell}},\ }\href {\doibase
  10.1088/1367-2630/ab4c8c} {\bibfield  {journal} {\bibinfo  {journal} {New
  Journal of Physics}\ }\textbf {\bibinfo {volume} {21}},\ \bibinfo {pages}
  {103048} (\bibinfo {year} {2019})}\BibitemShut {NoStop}%
\bibitem [{\citenamefont {Coulamy}\ \emph {et~al.}(2016)\citenamefont
  {Coulamy}, \citenamefont {Santos}, \citenamefont {Hen},\ and\ \citenamefont
  {Sarandy}}]{Coulamy}%
  \BibitemOpen
  \bibfield  {author} {\bibinfo {author} {\bibfnamefont {I.~B.}\ \bibnamefont
  {Coulamy}}, \bibinfo {author} {\bibfnamefont {A.~C.}\ \bibnamefont {Santos}},
  \bibinfo {author} {\bibfnamefont {I.}~\bibnamefont {Hen}}, \ and\ \bibinfo
  {author} {\bibfnamefont {M.~S.}\ \bibnamefont {Sarandy}},\ }\href {\doibase
  10.3389/fict.2016.00019} {\bibfield  {journal} {\bibinfo  {journal}
  {Frontiers in ICT}\ }\textbf {\bibinfo {volume} {3}} (\bibinfo {year}
  {2016}),\ 10.3389/fict.2016.00019}\BibitemShut {NoStop}%
\bibitem [{\citenamefont {Lu}\ \emph {et~al.}(2007)\citenamefont {Lu},
  \citenamefont {Miao},\ and\ \citenamefont {Metcalf}}]{MetcalfNonAdiabatic}%
  \BibitemOpen
  \bibfield  {author} {\bibinfo {author} {\bibfnamefont {T.}~\bibnamefont
  {Lu}}, \bibinfo {author} {\bibfnamefont {X.}~\bibnamefont {Miao}}, \ and\
  \bibinfo {author} {\bibfnamefont {H.}~\bibnamefont {Metcalf}},\ }\href
  {\doibase 10.1103/PhysRevA.75.063422} {\bibfield  {journal} {\bibinfo
  {journal} {Phys. Rev. A}\ }\textbf {\bibinfo {volume} {75}},\ \bibinfo
  {pages} {063422} (\bibinfo {year} {2007})}\BibitemShut {NoStop}%
\bibitem [{\citenamefont {Gao}\ \emph {et~al.}(1991)\citenamefont {Gao},
  \citenamefont {Xu},\ and\ \citenamefont {Qian}}]{Gao}%
  \BibitemOpen
  \bibfield  {author} {\bibinfo {author} {\bibfnamefont {X.-C.}\ \bibnamefont
  {Gao}}, \bibinfo {author} {\bibfnamefont {J.-B.}\ \bibnamefont {Xu}}, \ and\
  \bibinfo {author} {\bibfnamefont {T.-Z.}\ \bibnamefont {Qian}},\ }\href
  {\doibase 10.1103/PhysRevA.44.7016} {\bibfield  {journal} {\bibinfo
  {journal} {Phys. Rev. A}\ }\textbf {\bibinfo {volume} {44}},\ \bibinfo
  {pages} {7016} (\bibinfo {year} {1991})}\BibitemShut {NoStop}%
\bibitem [{\citenamefont {Kim}\ \emph {et~al.}(2000)\citenamefont {Kim},
  \citenamefont {Santana},\ and\ \citenamefont {Khanna}}]{Kim}%
  \BibitemOpen
  \bibfield  {author} {\bibinfo {author} {\bibfnamefont {S.}~\bibnamefont
  {Kim}}, \bibinfo {author} {\bibfnamefont {A.}~\bibnamefont {Santana}}, \ and\
  \bibinfo {author} {\bibfnamefont {F.}~\bibnamefont {Khanna}},\ }\href
  {\doibase https://doi.org/10.1016/S0375-9601(00)00406-0} {\bibfield
  {journal} {\bibinfo  {journal} {Physics Letters A}\ }\textbf {\bibinfo
  {volume} {272}},\ \bibinfo {pages} {46 } (\bibinfo {year}
  {2000})}\BibitemShut {NoStop}%
\bibitem [{\citenamefont {de~Ponte}\ \emph {et~al.}(2018)\citenamefont
  {de~Ponte}, \citenamefont {C\^onsoli},\ and\ \citenamefont
  {Moussa}}]{dePonte}%
  \BibitemOpen
  \bibfield  {author} {\bibinfo {author} {\bibfnamefont {M.~A.}\ \bibnamefont
  {de~Ponte}}, \bibinfo {author} {\bibfnamefont {P.~M.}\ \bibnamefont
  {C\^onsoli}}, \ and\ \bibinfo {author} {\bibfnamefont {M.~H.~Y.}\
  \bibnamefont {Moussa}},\ }\href {\doibase 10.1103/PhysRevA.98.032102}
  {\bibfield  {journal} {\bibinfo  {journal} {Phys. Rev. A}\ }\textbf {\bibinfo
  {volume} {98}},\ \bibinfo {pages} {032102} (\bibinfo {year}
  {2018})}\BibitemShut {NoStop}%
\end{thebibliography}%
\appendix

\section{Construction of the auxiliary angles and their boundary conditions from the invariant operator}
\label{invariantConstruction}

The construction of the invariant operator and its associated eigenvalue equation can be a difficult process, and various methods have been introduced to overcome this \cite{LRI,Gao,Kim,dePonte}. 
However, since we have the form of the Hamiltonian and its eigenvectors, we need only to parameterize $\hat I (t)$ and $\ket{\phi_n (t)}$ in the same functional forms.
Thus, in an identical manner to \cite{LRIandTQD}, we use the inverse engineering approach to parameterize $\delta (t)$ and $\Omega(t)$ as to begin and end the sweep with the desired populations. (Note that this derivation holds for time-dependent $\Omega(t)$ unless noted otherwise.) From Eqns.~\eqref{eigenvectors}, it follows that the invariant can be written in the basis of the Hamiltonian as
%It is clear that $\ket{\phi_\pm(t)}$ are eigenvectors of the invariant \cite{LRIandTQD}
\begin{equation}
	\hat I(t) = \frac{\hbar \Omega'}{2} \begin{pmatrix}
\cos \gamma &\sin \gamma \, e^{i \beta} \\
\sin \gamma \, e^{-i \beta} &- \cos \gamma
	\end{pmatrix}
\end{equation}
with eigenvalues $\lambda_{\pm} = \hbar \Omega' / 2$, where $\Omega'$ is an arbitrary constant frequency in order to keep $\hat I (t)$ with units of energy.
Substituting Eqns.~\eqref{eq:effectiveHam} and \eqref{eigenvectors} into Eq.~\eqref{LRPhases}, we calculate the Lewis-Riesenfeld phases for transfer 1 as 
\begin{equation}
\alpha_{\pm} (t) = \mp \frac{1}{2} \int_0^t \left( \delta \cos \gamma 
+  \Omega_1 \sin \gamma \cos \beta \right) dt' \text{.}
\end{equation}
Substituting these phases into Eq.~\eqref{LRIHamiltonian}, we find that the Hamiltonian is parameterized by 
\begin{equation} \label{ParameterizedH_R}
	\hat H_1^{(W)} (t) = \frac{\hbar}{2} \begin{pmatrix}
A &B e^{i \beta} \\
B^* e^{-i \beta} & -A
	\end{pmatrix} \text{,}
\end{equation}
where
\begin{align}
A &= \delta(t) \cos^2 \gamma + \Omega_1(t) \cos \gamma \sin \gamma \cos \beta \text{,} \\
B &= \delta(t) \cos \gamma \sin \gamma + \Omega_1(t) \sin^2 \gamma \cos \beta -i \dot \gamma \text{.}
\end{align}

Equating the two forms of the Hamiltonian [Eqns.~\eqref{eq:effectiveHam} and~\eqref{ParameterizedH_R}], we arrive at the auxiliary equations 
\begin{align} \label{AuxEq1App}
\dot \gamma &= \Omega_1(t) \sin  \beta  \text{,} \\
\label{AuxEq2App}
\delta & = \Omega_1(t) \cot \gamma \cos \beta \text{.}
\end{align}
The solutions to Eqns.~(\ref{AuxEq1App}) and~(\ref{AuxEq2App}) for~$\Omega_1(t)=\Omega=\text{const}$ is provided in the main text.

Next, we determine the boundary conditions on the auxiliary variables $\gamma$ and $\beta$ required for state transfer from state $\ket{G}$ to $\ket{E}$. As seen from Eq.~\eqref{eigenvectors}, $\gamma$ must satisfy 
\begin{equation}
\label{eq:gammaBoundaryApp}
    \gamma(t_0) = \pi n; \qquad
    \gamma(t_f) = \gamma(t_0)+ \pi(2m+1),
\end{equation} where $n$ and $m$ are integers. For simplicity, we choose $n=m=0$ so that $\gamma$ evolves from $0$ to $\pi$. Since the commutator between $\hat H_1^{(W)} (t)$ and $\hat I (t)$ is
\begin{equation}
	\begin{aligned}
\left[ \hat H_1^{(W)} (t), \hat I(t) \right] = \\
\frac{\hbar^2 \Omega'}{2} \Bigl( &
-i \Omega \sin \gamma \sin \beta \hat \sigma_W^z \\
&+ (\delta \sin \gamma e^{i \beta} - \Omega \cos \gamma )  \hat \sigma_W^{\dagger} \\
&+ (\Omega \cos \gamma - \delta \sin \gamma e^{-i \beta}) \hat \sigma_W \Bigr) \text{,}
	\end{aligned}
\end{equation}
where $\hat \sigma_W \equiv \ket{G}\bra{E}$, we should also impose
\begin{align}
\label{comm1App}
& \Omega(t^*) \sin \gamma (t^*) \sin \beta = 0 \text{,} \\
\label{comm2App}
&
\beta = q \pi
\end{align}
for $t^* = t_0$ and $t^*=t_f$ and integer $q$ to align the eigenbases of $\hat H_1^{(W)} (t)$ and $\hat{I}(t)$ at the beginning and end of the shortcut process (see Eq.~\eqref{eq:commutator}). The condition~\eqref{comm1App} is automatically satisfied by \eqref{eq:gammaBoundaryApp}, whereas the condition~\eqref{comm2App} is not necessary to enforce in the case of complete state transfer because it only affects non-physical, global phases.
%is usually enforced by using a time-dependent Rabi frequency that satisfies $\Omega(t_0)=\Omega(t_f)=0$. Nevertheless, we find from our numerical simulations with $\Omega = \text{const}$ that satisfying~\eqref{comm2App}, which reduces to $\beta=q \pi$ for integer $q$, is not necessary for our purposes.

We have now parameterized the Hamiltonian in terms of the auxiliary angles and found boundary conditions for these angles for the interaction with laser 1. The corresponding quantities for the subsequent interaction with laser 2 are derived in a similar manner.

\end{document}